\documentclass[fleqn,usenatbib]{mnras}

\usepackage{newtxtext,newtxmath}
\usepackage{lineno}
\usepackage[T1]{fontenc}

\DeclareRobustCommand{\VAN}[3]{#2}
\let\VANthebibliography\thebibliography
\def\thebibliography{\DeclareRobustCommand{\VAN}[3]{##3}\VANthebibliography}

\usepackage{graphicx}	
\usepackage{amsmath}	

\newcommand{\GALFORM}{{\tt GALFORM\,}}
\newcommand{\getemlines}{{\tt GET\_EMLINES}\,}

\title[Emission lines]{Modelling emission lines in star forming galaxies}

\author[C. M. Baugh et al.]{
C. M. Baugh$^{1}$\thanks{E-mail: c.m.baugh@durham.ac.uk (CMB)},
Cedric G. Lacey$^{1}$,
Violeta Gonzalez-Perez$^{2,3}$
and Giorgio Manzoni$^{1}$
\\
$^{1}$Institute for Computational Cosmology, Department of Physics, Science Laboratories, Durham University, \\
South Road, Durham, DH1 3LE, UK\\
$^{2}$Departamento de F\'isica Te\'orica, Facultad de Ciencias, Universidad Aut\'onoma de Madrid, ES-28049 Madrid, Spain\\
$^{3}$Centro de Investigaci\'on Avanzada en F\'isica Fundamental (CIAFF), Facultad de Ciencias, Universidad Aut\'onoma de Madrid, ES-28049 Madrid, Spain
}

\date{Accepted XXX. Received YYY; in original form ZZZ}

\pubyear{2020}

\begin{document}
\label{firstpage}
\pagerange{\pageref{firstpage}--\pageref{lastpage}}
\maketitle

\begin{abstract}
We present a new model to compute the luminosity of emission lines in star forming galaxies and apply this in the semi-analytical galaxy formation code \GALFORM. The model combines a pre-computed grid of HII region models with an empirical determination of how the properties of HII regions depend on the macroscopic properties of galaxies based on observations of local galaxies. The new model gives a very good reproduction of the locus of star-forming galaxies on standard line ratio diagnostic diagrams. The new model shows evolution in the locus of star forming galaxies with redshift on this line ratio diagram, with a good match to the observed line ratios at $z=1.6$. The model galaxies at high redshift have gas densities and ionisation parameters that are predicted to be $\approx 2-3$ times higher than in local star forming galaxies, which is partly driven by the changing selection with redshift to mimic the observational selection. Our results suggest that the observed evolution in emission line ratios requires other HII region properties to evolve with redshift,  such as the gas density, and cannot be reproduced by HII model grids that only allow the gas metallicity and ionisation parameter to vary.
\end{abstract}




\section{Introduction}

The energetic photons produced predominantly by hot, young stars can ionise the gas close to them, making   HII regions. The nebular emission which results as electrons and nuclei recombine or atoms and ions decay from collisionally excited states adds both continuum and emission line components to the composite stellar spectrum of a galaxy (see e.g. \citealt{Charlot:2001}, \citealt{Byler:2017}; for a review of emission lines see  \citealt{Kewley:2019}). The luminosity of emission lines depends on several factors, such as the rate of production of Lyman continuum photons which can ionise Hydrogen, thus probing the very recent star formation history in a galaxy, the metallicity of the star-forming gas and the density of the gas clouds in which young stars form. 

The ratios of various emission line luminosities  can be used to deduce the physical properties of HII regions. The classic emission line diagnostic diagram, introduced by \cite{Baldwin:1981} and referred to as the BPT diagram, shows two ratios of line luminosities, [OIII]/H$\beta$, plotted against [NII]/H$\alpha$. As the wavelengths of the emission lines in each ratio are close to one another, they will be affected similarly by dust attenuation, so in principle extinction does not change the value of the ratio. Also, since each emission line luminosity is proportional to the rate of production of Lyman continuum photons, the ratio of line luminosities does not  depend on this quantity in a simple, direct way (the rate of production of Lyman continuum photons can change the ionisation state of the HII region and hence the line ratios; see later).  Star forming galaxies occupy a distinct locus on the BPT diagram. The position of a galaxy on the BPT diagram can be used to distinguish between galaxies in which the emission lines are powered by star formation and those in which the lines are characteristic of the harder ionising spectrum associated with an active galactic nucleus \citep{Kewley:2001,Kauffmann:2003,Brinchmann:2004,Kewley:2008}. In principle, the position of a galaxy on the BPT diagram can be used, along with an HII region model, to infer the ionisation state of the typical HII region in a galaxy and the metallicity of the star-forming gas, once assumptions have been made regarding other properties of the HII regions, such as the ionised gas density. As we will see, there is some degeneracy in the information encoded in these line ratios, depending on which part of the BPT diagram the galaxy occupies and the particular HII region model used.   

The locus occupied by star forming galaxies in the BPT diagram is observed to evolve with redshift    \citep{Erb:2006,Steidel:2014,Kashino:2017,Strom:2017,Kashino:2019a}. The median [OIII]/H$\beta$ line ratio is measured to be $0.1$--$0.5$ dex higher at $z\sim 2.3$ than it is at $z=0$. A number of studies have argued that this increase in the [OIII]/H$\beta$ line ratio requires HII regions with a higher ionisation parameter \citep{Kewley:2001,Steidel:2014,Bian:2020}. 
\cite{Kewley:2001} modelled the emission line ratios in a sample of intense local starbursts that occupy the same region of the BPT diagram as high-redshift galaxies. Kewley et~al. found that the line ratios observed in these starbursts could only be reproduced when using stellar population models with hard spectra, due to their treatment of Wolf-Rayet stars; this led to the `maximum starburst line' that we will see plotted later on the BPT diagram.  \cite{Steidel:2014} also argued that the line ratios found in high redshift galaxies were characteristic of harder ionising radiation fields, similar to those in stellar population models that include the effects of binary stars \citep{Stanway:2014,Xiao:2018} or stellar rotation, which can affect the number and  type of Wolf-Rayet stars produced, which in turn changes the hardness of the ionising spectrum \citep{Conroy:2013}\footnote{We note that a top-heavy stellar initial mass function would also lead to a harder ionising spectrum; \cite{Kewley:2013} remark on such a possibility but decided that the evidence in favour of top-heavy IMF at the time was not strong enough to investigate such models further.}.
\cite{Bian:2020} reached a similar conclusion based on their analysis of the local counterparts of high redshift galaxies. 

The evolution of the locus of star forming galaxies on the BPT diagram was established using relatively small samples numbering a few hundred \citep{Strom:2017} to a couple of thousand galaxies \citep{Kashino:2019a}. Upcoming surveys will bring a dramatic increase in the data available on emission line galaxies at all redshifts. The Multi-Object Optical and Near-infrared Spectrograph (MOONS) under construction for the Very Large Telescope will be used to compile a sample of several hundred thousand galaxies in the redshift range $z=0.9-2.6$, with  spectra measured at a resolution of $R\sim 7000$, allowing accurate measurement of emission line luminosities  \citep{Maiolino:2020}. The Webb Space Telescope will track the classic BPT diagram emission lines as they are shifted into the infra-red at $z>5$, and also look at a wider variety of line ratios to probe conditions in the interstellar medium at different redshifts \citep{Gardner:2006}. 
Wide field surveys will use emission line galaxies to probe the large-scale structure of the Universe around $z \sim 1$ (e.g. the Dark Energy Spectroscopic Instrument survey aims to measure more than 20 million emission line galaxy redshifts, many of them OII emitters, see  \citealt{Raichoor:2020}). These observations mean that we will have information about the abundance, clustering and other properties of emission line galaxies, over a broad period of cosmic history covering the peak in the global star formation activity. 

This makes it timely to revisit the predictions for emission line galaxies in theoretical models. Some lines, such as H$\alpha$, are relatively straightforward to predict from the star formation rate \citep{Kennicutt:1998}, with a conversion factor that is almost independent of the metallicity of the star-forming gas and its ionisation parameter. Other lines, such as OII, are more complicated to predict and require a HII region model to be included in the galaxy formation model (see e.g. \citealt{Favole:2019ouk, Knebe:2021}). A further consideration is the attenuation of the emission line luminosity due to dust, which may be stronger than it is for the stellar continuum at the same wavelength. 

A model for the generation of emission lines from interstellar gas that is photoionized by young stars was implemented in the \GALFORM semi-analytical model of galaxy formation from its inception \citep{Cole:2000}. The model uses the tabulated output from the HII models of \cite{Stas:1990} for a given electron density, number of ionising stars per HII region and stellar effective temperature. The remaining variable after these choices have been made is the metallicity of the star forming gas, which is assumed to be the same as the global cold gas metallicity in the disk or bulge. This model has been used to make predictions for Lyman-$\alpha$ emitters by  \cite{LeDelliou:2005,LeDelliou:2006,Orsi:2008}, by \cite{Violeta:2018} to make predictions for the luminosity function and halo occupation distribution of OII emitters and by \cite{Violeta:2020} to establish the location of these galaxies in the cosmic web of large-scale structure. As we show in Appendix A, the original \GALFORM  emission line model gives a poor reproduction of the local BPT diagram. \cite{Orsi:2014} implemented the {\tt GET\_EMLINES} emission line model, based on the MAPPINGS-III photo-ionisation code results of \cite{Levesque:2010}, into the SAGS semi-analytical model of \cite{Gargiulo:2015}. Orsi et~al. reduced the HII model grid from two parameters, gas metallicity and ionisation parameter, to one by making the ionisation parameter a function of the gas metallicity. This model reproduces the local BPT diagram, but does not evolve with redshift (see Appendix B). This model has also been used to look at the clustering of emission line galaxies in different semi-analytical models \citep{Izq:2019,Favole:2019ouk}; a similar approach has also been applied in gas dynamic simulations \citep{Shen:2020}.  

\cite{Hirschmann:2017} used the HII region models of \cite{Gutkin:2016} to post-process 20 galaxies drawn from a series of cosmological zoom hydrodynamical simulations (see also \citealt{Hirschmann:2019}). Hirschmann et~al. included the contribution to emission lines from young stars, accretion onto supermassive black holes and older stellar populations that produce ionising photons (post asymptotic giant branch stars). With this sample of galaxies, Hirschmann et~al. were able to reproduce the local BPT diagram and its evolution for star forming galaxies. \cite{Zhai:2019} also coupled their galaxy formation model with the {\tt CLOUDY} model to compute emission lines.  

Here, we introduce a new model for star forming emission line galaxies that has similarities to that introduced by \cite{Hirschmann:2017}. We also use the HII region models of \cite{Gutkin:2016}, although these could be readily exchanged for other HII models. 
We use a different model to connect the global properties of the star forming gas in a \GALFORM galaxy to the properties of an HII region than was used by Hirschmann et~al. We do not consider emission lines resulting from accretion onto supermassive black holes, leaving this for future work. However, we are able to explore a much wider range of galaxy masses and host halo masses; the N-body simulation in which \GALFORM is implemented allows us to follow galaxy formation in halo masses covering more than  five decades in mass, rather than the single decade that Hirschmann et~al. were able to follow in their zoom resimulations.

The structure of the paper is as follows. In Section 2 we provide the theoretical background. This starts with a brief overview of the \GALFORM galaxy formation model (\S~2.1), followed by a description of the HII region model used in \S~2.2. In \S~2.3 we explain how the properties of the star forming gas predicted by \GALFORM are connected to the HII region model parameters. In Section 3 we present our main results, showing the model predictions for the BPT diagram at low redshift (\S~3.1), and intermediate (\S~3.2) and high redshift (\S~3.3). We discuss the change in the properties of the HII regions with redshift in \S3.4. We present a summary and our conclusions in \S~4. Some other emission line models are discussed in  Appendices A, B and C. The main paper focuses on the line ratios in the classic BPT diagram; predictions for other commonly used line ratios are compared with observations in Appendix D. 

\section{Theory}

Here we describe the three components of our new model for emission line galaxies: 1) the galaxy formation model (\S2.1), 2) the model of HII regions (\S2.2) and 3) the model for relating the properties of HII regions to macroscopic galaxy properties (\S2.3). This is a new treatment of emission lines in \GALFORM. To allow comparison with published work, a description of the previous HII region model used in \GALFORM is given in Appendix A; the {GET\_EMLINES} model of \cite{Orsi:2014} that has been used in other semi-analytical galaxy formation models is discussed in Appendix B. In Appendix C we show the variation in the grids of emission line luminosities returned by two current HII region models.

\subsection{Galaxy formation model} 

We use the \GALFORM semi-analytical model of galaxy formation \citep{Cole:2000,Bower:2006,Lacey:2016}. The model follows the key physical process that shape the formation and evolution of galaxies in the cold dark matter cosmology (for reviews of these processes and semi-analytical models see \citealt{Baugh:2006} and \citealt{Benson:2010}). In particular, the model tracks the transfer of mass and metals between different reservoirs of baryons, predicting the chemical evolution of the gas that is available to form stars and the full star formation history of galaxies. Two modes of star formation are considered: quiescent star formation in disks and bursts of star formation triggered by mergers or the motion of gas driven by dynamically unstable disks. The model  combines this information with a simple stellar population synthesis model to predict the luminosity of galaxies in different bands, and, importantly for emission lines, the number of Lyman continuum photons\footnote{Calculated using e.g. eqn. 4 from \cite{Gutkin:2016}.} that are available to change the ionisation state of hydrogen (see, for example, \citealt{Violeta:2014}). 

Here we use the versions of the model introduced by \cite{Violeta:2014}  (hereafter GP14) and \cite{Lacey:2016} (hereafter L16),  as recalibrated by \cite{Baugh:2019} following their implementation in the P-Millennium N-body simulation. 
The main difference between the two recalibrated models that is of relevance here is the form of the stellar initial mass function (IMF) adopted in different modes of star formation. In the \cite{Violeta:2014} model, a solar neighbourhood IMF is assumed in both quiescent star formation in galactic disks and bursts triggered by mergers or dynamically unstable disks. In the \cite{Lacey:2016} model, star formation in bursts is instead assumed to take place with a top-heavy IMF. This variation of the IMF was introduced by \cite{Baugh:2005} and allowed the model to reproduce, at the same time, the number counts and redshift distribution of dusty star forming galaxies detected by their emission at sub-millimetre wavelengths, and the properties of local galaxies in the optical and near-infrared (see also \citealt{Lacey:2008,Lacey:2016}). The model with a top-heavy IMF also matches the evolution of the galaxy luminosity function in the ultraviolet to high redshift \citep{Lacey:2011}. Whilst the need for a top-heavy IMF is disputed by some authors (see e.g. \citealt{Hayward:2013,Lagos:2019}; for observational evidence in support of a top-heavy IMF see the recent summary in  \citealt{Cowley:2019}), this choice gives the model some interesting phenomenology: i) 
the yield of metals in bursts is twice that in quiescent star formation for the top-heavy IMF which is adopted, with slope $x=1$ (the solar neighbourhood IMF used in quiescent star formation has a slope of $x=1.5$ above one solar mass; for reference, the \cite{Salpeter:1955} IMF has a slope of $x=1.35$ in this convention). ii) for a given star formation rate, bursts with an $x=1$ IMF produce three times as many Lyman continuum photons as for a solar 
neighbourhood IMF. Overall, quiescent star formation dominates at low redshift and burst star formation becomes more important with increasing redshift (see figure~26 of \citealt{Lacey:2016}). Note that the rate of production  of Lyman continuum photons affects the model predictions for the BPT diagrams, even though these are line ratios, because we need to consider the disk and bulge components of each galaxy individually, and these can have different amounts of star formation. Hence the number of Lyman continuum photons does not cancel out trivially if both the disk and bulge have ongoing star formation. The number of Lyman continuum photons will affect the ionisation parameter used in the HII region model (see later) and does influence the model predictions for the luminosity functions of emission line galaxies. Also, the use of a top-heavy IMF will lead to a harder ionising spectrum which could alter the conditions in the HII regions; we do not address this here, but discuss this as future work in the Conclusions.

\subsection{Nebular emission model}

We use the pre-computed grid of emission line luminosities released by \cite{Gutkin:2016}\footnote{http://www.iap.fr/neogal/models.html}. \cite{Gutkin:2016} calculated the grid by combining the \cite{BC:2003} stellar population synthesis model with the c13.03 release of the {\tt CLOUDY} photoionisation code  \citep{Ferland:2013}, following the methodology set out by \cite{Charlot:2001}. Gutkin et~al. assume that the HII regions have a spherical geometry, with gas at density $n_{H}$ occupying a fraction, $\epsilon$, of the volume within the radius of the HII region, given by the Str\"{o}mgren radius $R_{\rm s}$. The HII regions expand as more ionising photons are produced.

There are three main parameters which describe the Gutkin et~al. HII region models: i) the metallicity of the cold gas in the interstellar medium, $Z$, ii) the ionisation parameter, $U$, (see below for a definition) and iii) the hydrogen gas density $n_{\rm H}$. The metallicity in the Gutkin et~al. grid varies from $Z = 10^{-4}$ to $0.04$\footnote{Note that the value assumed by \cite{Gutkin:2016} for solar metallicity is $Z_{\odot}=0.01524$.}. The HII regions are considered to be ionisation-bounded i.e. the extent  of the HII region depends on the supply of ionising photons which is smaller than the total number of hydrogen atoms in the gas cloud. The edge of the HII region, given by the Str\"{o}mgren radius, $R_{\rm s}$, lies within the edge of the cloud of gas surrounding the hot, young stars. The dimensionless ionisation parameter is measured at the Str\"{o}mgren radius and is given by 

\begin{equation}
   U(t, R_{\rm s}) = \frac{ Q(t)}{4 \pi R^{2}_{\rm s} n_{\rm H} c}, 
\end{equation}
where $Q(t)$ is the rate of production of ionising photons i.e. photons with a wavelength less than 912 \AA. The ionisation parameter is usually quoted for a zero-age stellar population, $U=U(t=0)$. The ionisation parameter values used in the Gutkin et~al. grid vary from $\log U = -4.0$ to $\log U = -1.0$. The density of the interstellar medium is quoted is terms of the hydrogen gas density, $n_{\rm H}$, since we are considering hydrogen ionising photons. We consider hydrogen gas densities of $n_{\rm H} = 10, 100$ and $1000\, {\rm cm}^{-3}$. For each value of $n_{\rm H}$, the grid of ionisation, $U$, and metallicity, $Z$, is available over the values stated. 
The availability of a contemporary pre-computed grid of HII models covering $U, Z$ {\it and} $n_{\rm H}$ is unusual, and potentially important for studying the evolution of the BPT diagram line ratios with redshift. 
The other HII model parameters are held fixed (see Table 3 of Gutkin et~al. for the full range of parameter values for which HII models are available): 
 the number weighted ${\rm C}/{\rm O}$ ratio is held at the solar value of $({\rm C}/{\rm O})_{\odot}=0.44$, the upper mass cut-off of the stellar initial mass function is set to $100\, M_{\odot}$, the dust-to-metal mass ratio, $\xi_{\rm d}$, is set to 0.3, which is slightly lower than the local interstellar medium value of 0.36, but for which a wider range of hydrogen gas density values is available (see Gutkin et~al.). We consider the impact of using a different HII region model grid in Appendix C.
 
\begin{figure}
\includegraphics[trim=0.75cm 1.5cm 0cm 0.5cm,clip,width=1.0\columnwidth]{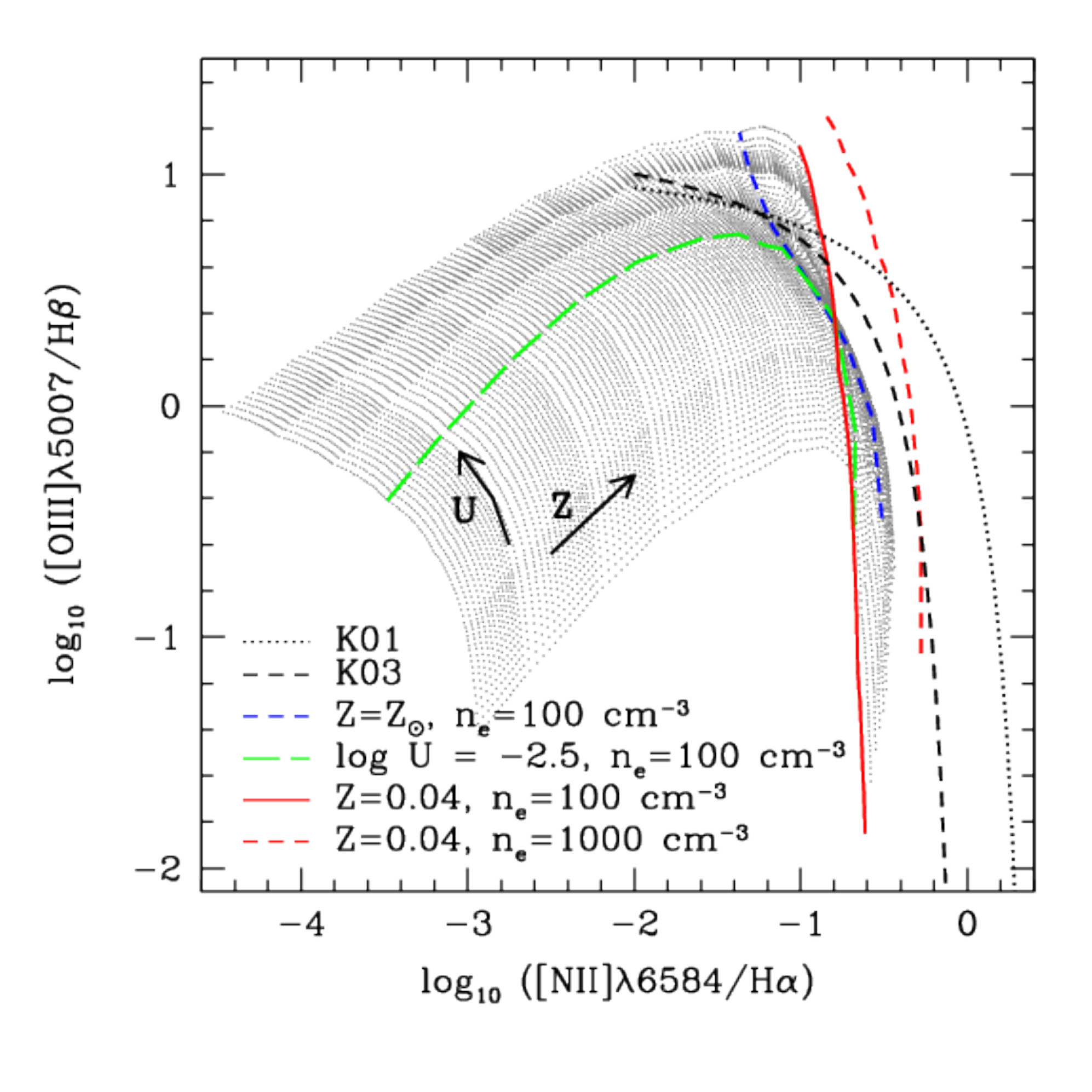}
    \caption{The pre-computed BPT diagram grid from Gutkin et~al. (2016). The grey points show the grid for a gas density of $n_{\rm H}=100 \, {\rm cm}^{-3}$, varying the ionisation parameter $U$ and metallicity $Z$ as indicated on  logarithmically spaced grids. The green curve shows the line ratios predicted for a fixed ionisation parameter of $\log U = -2.5$, varying $Z$, and the blue curve shows the case of fixed solar metallicity, varying $U$. The highest metallicity value of this grid is marked by the solid red line. This maximum metallicity line shifts to the dashed red line on increasing the gas density to $n_{\rm H} = 1000 \, {\rm cm}^{-3}$. The black dashed line shows the relation separating emission powered by star formation from that driven by AGN, as estimated by Kauffmann et~al. (2003). The black dotted curve shows the maximum starburst line from Kewley et~al.(2001). See text for more details. }
    \label{fig:bpt_gutkin}
\end{figure}

The pre-computed grid from \cite{Gutkin:2016} is shown in Fig.~\ref{fig:bpt_gutkin} for the classic BPT diagram line ratios, plotting the ratio of the line luminosities of [NII]/H$\alpha$ on the $x$-axis and [OIII]/H$\beta$ on the $y$-axis. The grey points show the grid of line ratios for a hydrogen gas density of $n_{\rm H}=100 \, {\rm cm}^{-3}$, sampled on a much finer grid of ionisation parameter and ISM metallicity than the published tables, using bilinear interpolation in $\log Z$ and $\log U$. The Gutkin et~al. grid extends to very low metallicities and hence lower [N{\sc II}]/H$\alpha$ ratios than are typically observed for star-forming galaxies. This can be appreciated by the location of the blue curve in Fig.~\ref{fig:bpt_gutkin}, which shows the line ratios predicted for stars forming from gas with solar metallicity, which is almost at the right hand edge (highest metallicity) of the grid. The green curve shows how the line ratios change with metallicity for a fixed ionisation parameter of $\log U = -2.5$. For most of the grid, the line ratios uniquely determine the values of $U$ and $Z$. This breaks down for gas metallicities close to solar, as the grid points fold over on themselves; the solid red line shows the ratios predicted for the highest metallicity available. For this hydrogen gas density, the highest values of the line ratio [NII]/H$\alpha$ do not always correspond to the highest metallicity (e.g. particularly at lower values of [OIII]/H$\beta$). Unfortunately, as we will see in \S~3.1, this is where most observed galaxies fall on the BPT diagram. 

Part of the motivation for using the pre-computed models from Gutkin et~al. is that grids of line luminosities are available for different hydrogen gas densities. As we will see later, there is some expectation that the conditions in the HII regions, such as the density, could change with redshift. At low metallicities, the emission line grids are very similar, independent of the value of $n_{\rm H}$. There is a shift in the model grid at high metallicities on moving to higher hydrogen gas densities. The model grid for the highest metallicity is highlighted by the red dashed line for $n_{\rm H} = 1000\, {\rm cm}^{-3}$ in Fig.~\ref{fig:bpt_gutkin}. For this hydrogen gas density, this line {\it does} mark the extreme values of [NII]/H$\alpha$ on the model grid, unlike the case for $n_{\rm H} = 100 \,{\rm cm}^{-3}$, where the grid folds back on itself and the highest values of [NII]/H$\alpha$ are not achieved for the highest HII region metallicity. There is a clear shift in the right hand boundary of the model grid on moving from $n_{\rm H} = 100 \,{\rm cm}^{-3}$ to $n_{\rm H} = 1000 \,{\rm cm}^{-3}$. Potentially, this will allow a model that uses this grid to reproduce the evolution observed with redshift in the line ratios for star-forming galaxies. We note that for the lowest gas density available in the Gutkin et~al. models, $n_{\rm H}=10\, {\rm cm}^{-3}$, there is only a modest shift in the line ratio values for the highest gas metallicity, which move slightly to the left of those indicated by the red solid line in Fig.~\ref{fig:bpt_gutkin}.

Fig.~\ref{fig:bpt_gutkin} also shows two relations which are often used to set the context for the discussion of the BPT diagram. The black dashed line is the classification boundary  introduced by \cite{Kauffmann:2003} to separate line ratios powered mostly by star formation (to the left of the line) from those in which AGN emission dominates (to the right of the line). This classification was inferred from observations of local galaxies and does not apply to high redshift samples, as we shall see later. The dotted black line is the `maximum starburst' line, a theoretical limit proposed by \cite{Kewley:2001}. Even though in both cases these relations are intended to indicate the maximum line ratios for emission lines powered by star formation, the Kewley et~al. curve is higher than the Kauffmann et~al. one because Kewley et~al adopted a stellar population synthesis model with a harder UV spectrum (due to the particular treatment used to model Wolf-Rayet stars; see \citealt{Dagostino:2019} for a comparison of the emission line properties using different population synthesis and photoionisation models). 

\subsection{Connecting the properties of HII regions to galaxy properties}

The parameters of the HII region models in \cite{Gutkin:2016} are viewed as `effective' parameters that apply to the ensemble of HII regions within a galaxy (see also the discussion in \citealt{Charlot:2001}). This is motivated by the similarity found between the optical emission line ratios observed in star forming galaxies and those measured in individual HII regions \citep{Kobulnicky:1999}. Due to this similarity, a model of an individual HII region can be used to represent the emission lines from a whole galaxy, provided that there is some means to connect the values of global properties measured in a galaxy to the conditions  within a HII region. 

Here we use an empirical determination of the connection between global galaxy properties and the properties within HII regions by \cite{Kashino:2019b}. These authors investigated  the correlation between two global galaxy properties (stellar mass, $M_*$, and specific star formation rate, $sSFR$) and three properties of HII regions ($U, Z, n_{\rm e}$), for local galaxies using spectra from the Sloan Digital Sky Survey \citep{SDSS:DR7}. The HII region properties were determined using various standard line ratios and HII region modelling, combining the Starburst99 stellar population synthesis code of \cite{Leitherer:1999} with {\tt CLOUDY} (version C17.00). Note that Kashino \& Inoue cast their results in terms of the density of electrons, $n_{\rm e}$, inside the HII region, which is close to the density of Hydrogen gas. We take their $n_{\rm e}$ as a proxy for $n_{\rm H}$ in the HII region models of Gutkin et~al. To avoid confusion when referring back to Kashino \& Inoue, we retain their $n_{\rm e}$ notation in the equations presented in this subsection. 

Interestingly, \cite{Kashino:2019b} found that the partial dependence of the ionisation parameter on metallicity, $U \propto Z^{-0.36}$, i.e. when allowing $U$ to depend on other properties, is much weaker than the apparent correlation, $U \propto Z^{-1.52}$, when only allowing $Z$ to vary. Kashino \& Inoue comment that this arises due to the anti-correlation they find between specific star formation rate and metallicty; both these quantities appear in the full fit for the ionisation parameter as we shall see below.  

We adopt the best fitting relations for the HII region parameters from Table 2 from \cite{Kashino:2019b} as follows: 

\begin{eqnarray}
\log_{10}\left[\frac{n_{\rm e}}{{\rm cm^{-3}}}\right] \hspace{-0.1cm} &=& 2.066 + 0.310  \left( \log_{10}(M_{*}/M_{\odot})-10.0 \right) \label{eq:ne}\\ \nonumber
&+& 0.492 \left(\log_{10}(sSFR^{\prime}/{\rm yr^{-1}}) + 9 \right) \\ 
\log_{10} U &=& - 2.316 - 0.36\left(0.69  
 + \log_{10}(Z_{\rm cold}/Z_{\odot}) \right) \label{ref:u} \\  \nonumber  
&-&0.292 \log_{10} (n_{\rm e}/{\rm cm}^{-3}) \\ \nonumber 
&+& 0.428 \left(\log_{10}(sSFR^{\prime}/{\rm yr^{-1}}) + 9\right), 
\end{eqnarray}
where $Z_{\rm cold}$ is the metallicity of the cold gas, $M_{*}$ is the stellar mass of the galaxy, $sSFR^{\prime}$ is the effective instantaneous specific star formation rate in units of ${\rm yr}^{-1}$ (the prime indicates that, as explained below, this quantity is calculated in a slightly different way than the standard specific star formation rate).  Kashino \& Inoue adopt a solar metallicity of  $Z_{\odot} = 0.014$ for the metallic mass
fraction, following \cite{Asplund:2009}. The above relations for $n_{\rm e}$ and $U$ are applied to the disk and bulge components of the model galaxies separately; the bulge component is only considered if there is an ongoing starburst. We use the values of $Z_{\rm cold}$ and $M_{*}$ for the disk and bulge output directly by \GALFORM. The specific star formation rate is computed from the rate of production of ionising photons output by the model to obtain an {\it effective} star formation rate that is relevant to emission lines. For the \cite{Kennicutt:1998} IMF, we assume that a constant star formation rate of $1 M_{\odot} {\rm yr}^{-1}$ produces Lyman continuum photons at a rate of $9.85 \times 10^{52}$ photons $\rm s^{-1}$. The top-heavy IMF with $x=1$, which is adopted in bursts of star formation which add stars to the galactic bulge, gives $30.83 \times 10^{52}$ photons ${\rm s}^{-1}$ for the same star formation rate.  If a model galaxy is predicted to have $N_{\rm Ly}$ photons per second associated with star formation in its disk or bulge, then we compute the effective star formation rate ($SFR^{\prime}$) from this as follows: $SFR^{\prime} = N_{\rm Ly}/(9.85 \times 10^{52}) M_{\odot} {\rm yr}^{-1}$, ie we use the conversion factor from the rate of production of Lyman continuum photons to star formation rate for a solar neighbourhood IMF. Hence for same actual star formation rate, star formation in a burst with a top-heavy IMF will lead to a higher effective star formation rate, as the rate of Lyman continuum photon production is higher in a burst than it is for quiescent star formation. The use of an effective star formation rate in this way matches more closely the star formation rate estimated by Kashino \& Inoue in their observational study.

We compute $n_{\rm e}$ first using Eqn.~2, and then use this to calculate the ionisation parameter, $U$, using Eqn.~3. As commented above we treat $n_{\rm e}$ as a proxy for the gas density, $n_{\rm H}$, in the HII region models, following the approach of Kashino \& Inoue, and retaining their notation in this section in an attempt to avoid confusion.

We apply the attenuation factor calculated in \GALFORM for the stellar continuum at the wavelength of the emission line. This calculation uses the size of the disk and bulge components of the galaxy computed by \GALFORM and the chemical evolution model implemented as described in \cite{Cole:2000}, \cite{Violeta:2013} and \cite{Lacey:2016}. Attenuation factors are computed separately for the disk and bulge components, and so these factors do not cancel out in the line ratios unless only one of these components is actively forming stars. 
We note that it is possible that, given their origin, emission lines could experience more attenuation than the stellar continuum (e.g. \citealt{Dominquez:2013}).

Our new model is straightforward. We use \GALFORM to predict the cold gas metallicity, stellar mass and rate of production of ionising photons. These quantities are available for the disk and bulge components of each galaxy. The stellar mass and rate of production of ionising photons are then used to calculate the specific star formation rate for the disk and bulge. With these properties, we can use Eqns.~{2} and {3} to calculate values of $U$ and $n_{\rm e}$ for the disk and bulge. These parameters, along with the respective $Z_{\rm cold}$, are then used to access emission line ratios from the pre-computed grid of HII models from \cite{Gutkin:2016}. We use bilinear interpolation in $\log Z$ and $\log U$ to extract line luminosities from the HII model grid. If the \GALFORM model values for $U, Z$ and $n_{\rm e}$ lie outside the range of the grid, then we use the limiting value from the grid that is closest to the \GALFORM prediction.  Following other studies of nebular emission (e.g. \citealt{Orsi:2014,Byler:2017}) we assume that all of the Lyman continuum photons produced contribute to the emission line luminosity; this assumption affects the luminosity of individual lines and could affect the line ratios in the BPT diagram by altering the ionisation parameter used in the HII region model. Observationally, Lyman continuum emission is seen in some galaxies with strong nebular emission \citep{Nakajima:2020ApJ}, though in general the escape fraction is considered to be low for emission line galaxies (e.g. \citealt{Rutkowski:2017} quote an escape fraction of less than 15 per cent at $z=2.5$).

\section{Comparison to observations} 

\begin{figure}
    	\includegraphics[trim=0.75cm 1.5cm 0cm 0.5cm,clip,width=1.0\columnwidth]{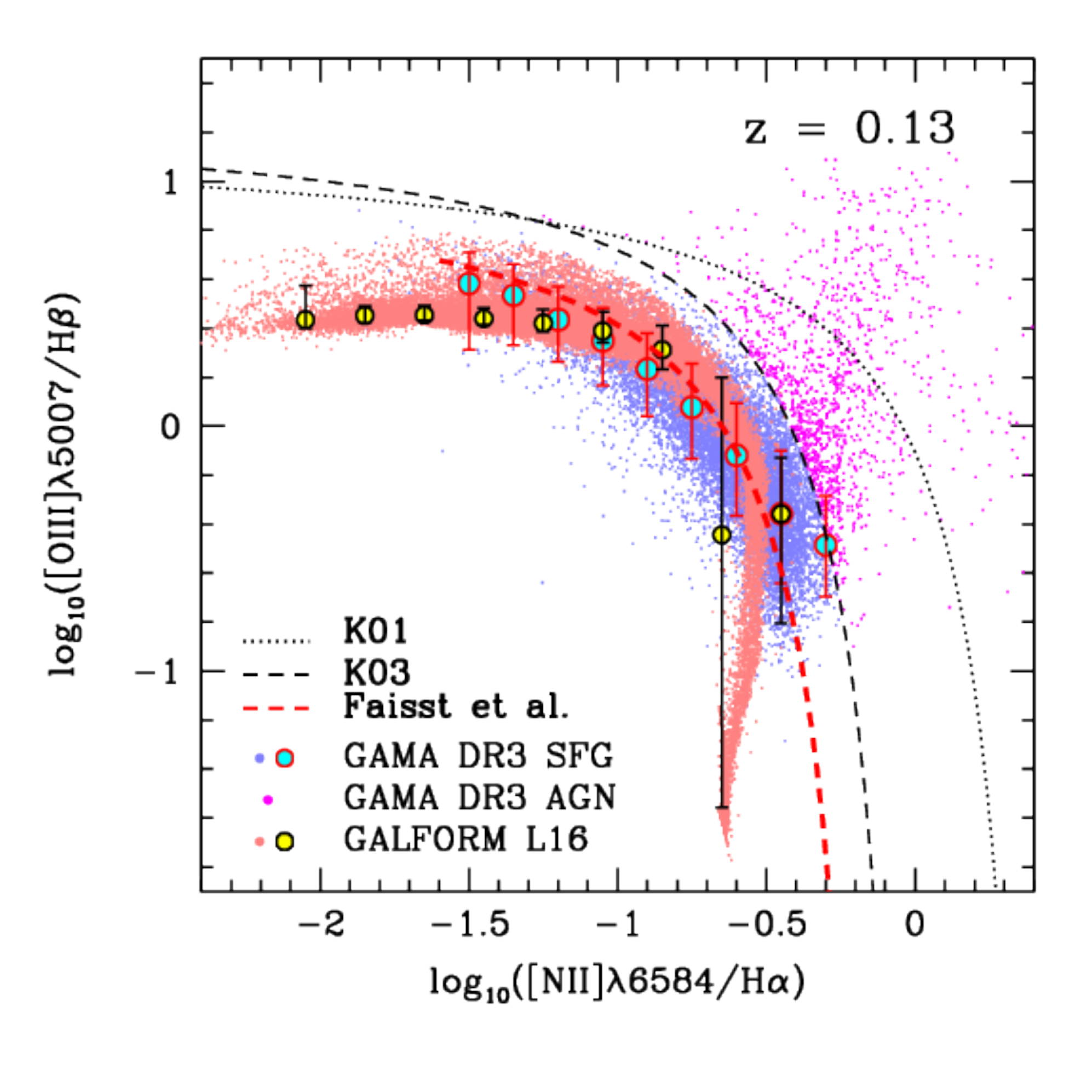}
    \caption{The BPT diagram at low redshift. The blue and magenta points show objects from GAMA DR3, selected as described in the text. 
    The colours are determined by which side of the black dashed line the points fall on: this line shows the division between star forming galaxies (blue) and active galactic nuclei (magenta) proposed by Kauffmann et~al. (2003). The black dotted line is the maximum starburst line from Kewley et~al. (2001). The dashed red line shows the fit from Faisst et~al. (2018) at $z=0.13$, the median redshift of the GAMA star forming galaxies. The large cyan (red edged) circles show the median line ratios for GAMA galaxies and the bars show the 10-90 percentile range. The median line ratios and 10-90 percentiles are shown for L16  \GALFORM  galaxies (yellow-filled circles) that meet the GAMA selection criteria. 
  }
    \label{fig:bpt_z0.131}
\end{figure}

Here we compare the new model to observations at low (\S~3.1) and high redshift (\S~3.2 \& \S~3.3). We probe further into the \GALFORM galaxy properties that match the approximate observational selection in each case in \S~3.4, to better understand the predicted evolution in the model BPT diagram.

 \subsection{Low redshift} 

We first compare the predictions of our new model for emission line galaxies to low redshift observations from the Galaxy And Mass Assembly survey Data Release 3 (GAMA DR3) \citep{Baldry:2018}. GAMA DR3 is made up of several fields which have slightly different magnitude limits. We focus on objects in the G15 and G02 fields which are the ones which have the deeper Petrosian magnitude limit of $r=19.8$ in DR3 (the other fields are complete to a brighter $r-$band flux in DR3). In the G02 region the spectroscopic completeness to this magnitude limit is highest north of $\delta = -6.0^\circ$, so we only consider this part of this region. The estimation of emission line properties from spectra is described in \cite{Gordon:2017}. In particular we use the GaussFitSimple table from the SpecLineSFR (v05) DMU (see Gordon et al. and the web link below for further details). We have checked that the observed BPT diagram is essentially unchanged if instead we use the DirectSummation table of emission line fluxes and equivalent widths. We restrict our  attention to objects with a redshift quality flag $nQ \ge 3$. Objects with a signal-to-noise ratio $S/N \ge 3$ in each of the emission lines which feature in the BPT diagram are retained. After downloading this dataset using the handy SQL query builder\footnote{http://www.gama-survey.org/dr3/query/qb.php}, the Balmer line fluxes are corrected for stellar absorption using the equivalent width, as set out in eqn.~5 from \cite{Gordon:2017}. 

We label objects as being star forming galaxies or AGN using the local classification line proposed by \cite{Kauffmann:2003}, as discussed in \S~2.2. 
GAMA DR3 objects below this curve, shown by the black dashed line in Fig.~\ref{fig:bpt_z0.131}, are considered to be star-forming galaxies and are plotted as blue points, and those above the line are assumed to be AGN, shown by magenta points, which we subsequently ignore. The locus of star-forming galaxies in the BPT plane is shown by the median [OIII]/H${\beta}$ ratio (blue points with red borders). The median redshift of these 10\,634 star forming galaxies is $z=0.130$. The red dashed line shows the fit to the locus of star-forming galaxies on the classic BPT diagram proposed by \cite{Faisst:2018}, evaluated at this redshift, which is in reasonable agreement with the medians from the GAMA sample. This agreement is noteworthy as the GAMA BPT diagram was not used to calibrate the Faisst et~al. fit.

The \GALFORM \, predictions are shown for galaxies at the closest simulation snapshot to the sample of GAMA star-forming galaxies, which is at $z=0.131$; the model galaxies have $r<19.8$ and an H${\alpha}$ flux greater than $2 \times 10^{-15}\, {\rm erg \, s}^{-1} {\rm cm^{-2}}$. The cut on line flux is an approximate comparison as we do not have noise in the theoretical predictions, and so it is hard to make a cut based on the signal-to-noise ratio of a line to match the selection applied to the observations. Inspection of the distribution of line fluxes in our sample of star-forming galaxies from GAMA DR3 shows a gradual decline in galaxy number with reducing flux rather than sharp cut in flux, as this is a multi-variate selection (i.e. involving the $r$-band magnitude and the flux in the other lines). The flux limit applied to the \GALFORM \, galaxies corresponds to the peak of the flux distribution in the GAMA sample  (note that the median H$\alpha$ flux of the selected GAMA star-forming galaxies is $2.79 \times 10^{-15}\, {\rm erg \, s}^{-1} {\rm cm^{-2}}$). Reassuringly, the model predictions are not that sensitive to the precise line flux cut applied. 

The predictions of the recalibrated GP14 and L16 models are quite similar; the median [OIII]/H$\beta$ line ratios in the L16 model galaxies are $\approx 0.05$ dex lower than those in the GP14 model. For this reason we only show the L16 model predictions in Fig.~\ref{fig:bpt_z0.131}. At higher values of [NII]/H${\alpha}$, the model predictions agree extremely well with the GAMA DR3 sample, except for the case of the bin centred on $\log$[NII]/H${\alpha} \approx -0.7$. For this bin, the median [OIII]/H${\beta}$ predicted by the model lies some way below the observed value. In fact, the model predictions are bimodal in this bin, as indicated by the points which show the line ratios for individual galaxies, which is why the errorbar showing the 10-90 percentile range is so broad. This disagreement is likely to be driven by how well we have been able to reproduce the observational selection. There is a clear edge apparent in the observed BPT diagram at low values of [OIII]/H${\beta}$; we will see in the next subsections that this lack of observed points is also apparent at high redshift.  If we rejected model galaxies that fall below this edge in the data, this would remove the clump of galaxies at very low values of [OIII]/H${\beta}$ (i.e. those with [OIII]/H$\beta < -0.1$), and bring the median line ratio predicted by the model into much better agreement with the observations in this bin. 
At lower values of [NII]/H${\alpha}$, the [OIII]/H${\beta}$ ratio does not rise as steeply in the model as suggested by the observations.

\subsection{Intermediate redshift} 

\begin{figure}   
    	\includegraphics[trim=0.75cm 1.5cm 0cm 0.5cm,clip,width=1.0\columnwidth]{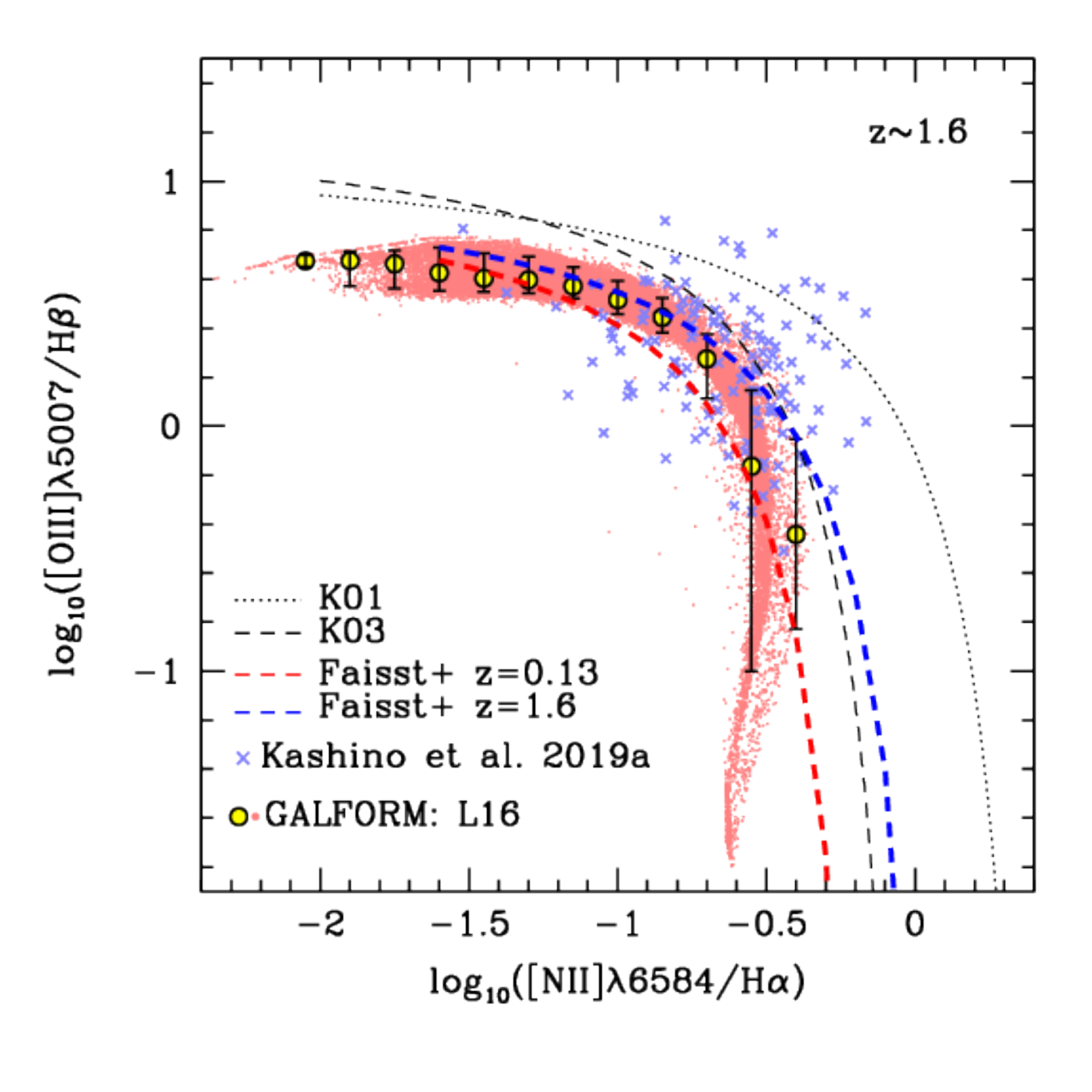}
    \caption{The BPT diagram at $z=1.6$. The blue crosses show galaxies from Kashino et~al. (2019a). 
    The blue dashed line shows the relation from Faisst et~al. (2018) for $z=1.6$, while the red line shows their fit at $z=0.131$ for reference. The circles show the median line ratios for the L16 \GALFORM galaxies meeting the sample selection (see text) and the bars show the 10-90 percentile range. The small faint red points show individual \GALFORM \, galaxies.} 
\label{fig:bpt_z1.6}
\end{figure}

We now compare the model predictions with a sample of intermediate redshift galaxies from the Fiber Multi-Object Spectrograph (FMOS)-COSMOS survey by \cite{Kashino:2019a} (see also \citealt{Silverman:2015,Kashino:2017}). \cite{Kashino:2017} used an earlier version of this dataset to confirm the offset in the location of intermediate redshift galaxies in the BPT plane compared with local galaxies that had previously been reported for an even higher redshift sample by \cite{Steidel:2014} (see \S~3.3).

Here we use the Release Version 2.0 provided by \cite{Kashino:2019a}\footnote{http://member.ipmu.jp/fmos-cosmos/FC\_catalogs.html}. Objects with $z$Flag=4 are retained, which means that the object's redshift has been confirmed using two emission lines, one with a $S/N \ge 5$ and the other $S/N \ge 3$. Following the approach taken by \cite{Kashino:2017} any objects that are either detected in the X-ray or which have a FWHM for H${\alpha}$ greater than
$1000 \, {\rm km \, s}^{-1}$ are discarded  as being a possible AGN. An object needs to be detected in H${\alpha}$ with $S/N>3$ and in all of the other lines in the BPT diagram with $S/N>1.5$. The flux limit in H${\alpha}$ is $4 \times 10^{-17} {\rm erg \, s}^{-1} \, {\rm cm}^{-2}$.

The resulting sample of $z\sim 1.6$ galaxies is plotted as blue crosses in Fig.~\ref{fig:bpt_z1.6}. Objects that lie above the maximum starburst line proposed by \cite{Kewley:2001} (black dotted line) are considered by \cite{Kashino:2019a}
to be AGN (but are still plotted in Fig.~\ref{fig:bpt_z1.6}). 
The fit to the locus of star-forming galaxies on the BPT diagram from \cite{Faisst:2018} for $z=1.6$ is shown by the blue dashed line. Note that the Kashino et~al. data were used to calibrate the Faisst et~al. fit.
The $z=0.13$ version of the Faisst et~al. curve is reproduced for reference to show the evolution in the observed BPT diagram. 

The \GALFORM predictions are shown in Fig.~\ref{fig:bpt_z1.6} as faint red points. The larger points with yellow cores show the median line ratios and the bars show the 10-90 percentile range of the model predictions. The model galaxies are at $z=1.6$ and have an H${\alpha}$ flux in excess of $4 \times 10^{-17} \, {\rm erg \, s}^{-1}\, {\rm cm}^{-2}$. 
There is good agreement between the model predictions and the FMOS observations, particularly for intermediate [NII]/H$\alpha$ ratios, with a clear shift to higher [OIII]/H${\beta}$ ratios beyond the red-dashed line which corresponds to the BPT locus at $z=0.131$.  The drop in the [OIII]/H${\beta}$ ratio seen in the model at high values of [NII]/H${\alpha}$ is likely to be the result of a selection effect that we are not imposing properly on the model, as discussed for the local BPT diagram. There are few observed galaxies with $\log$[OIII]/H$\beta < -0.5$, where as the model galaxies extend down to $\log$[OIII]/H$\beta \sim -1.5$

\subsection{High redshift}

\begin{figure}
    	\includegraphics[trim=0.75cm 1.5cm 0cm 0.5cm,clip,width=1.0\columnwidth]{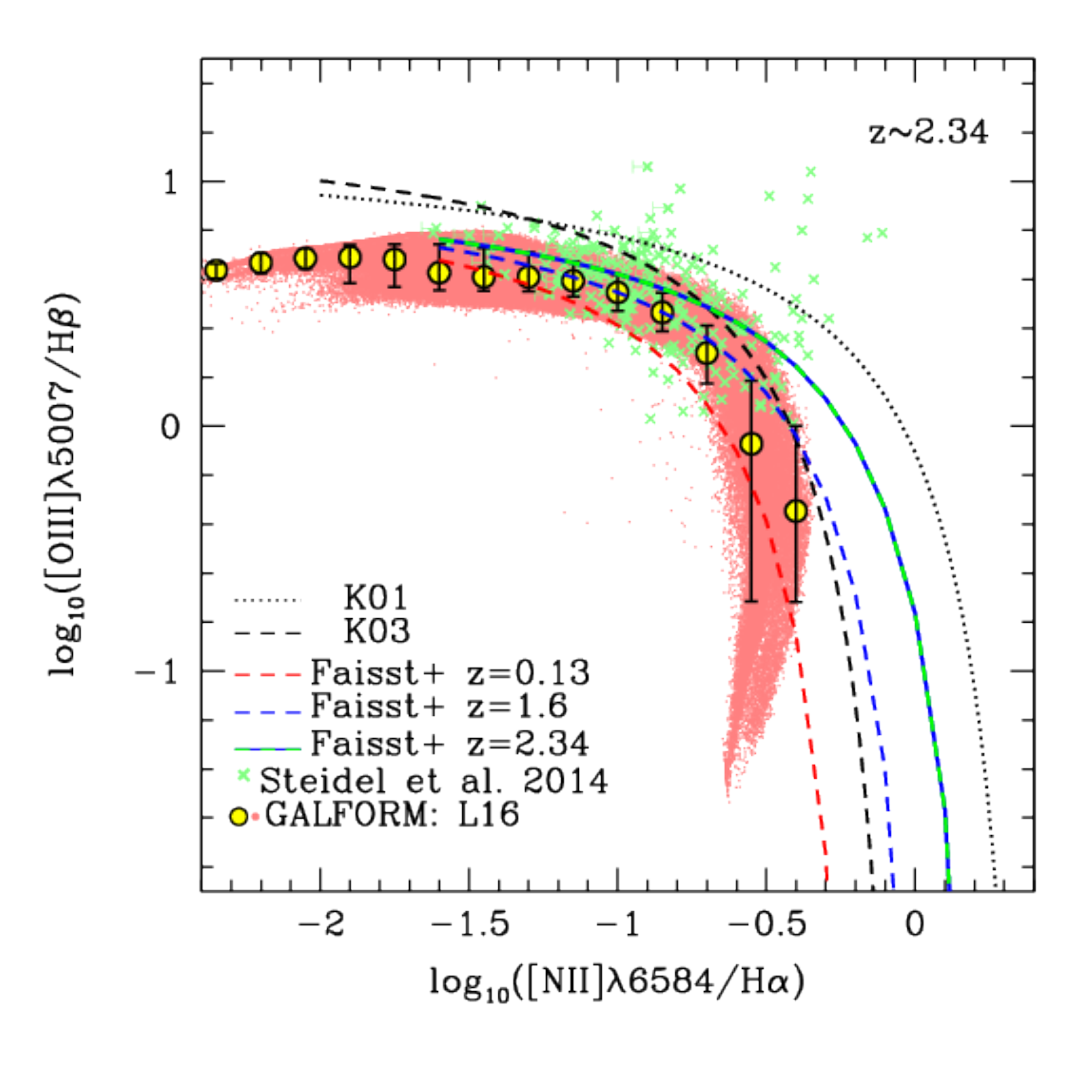}
    \caption{The BPT diagram at $z=2.3$. The green points show observational data from Steidel et~al. (2014). The green-blue dashed line shows the fit from Faisst et~al. (2018); the $z=0.131$ and $z=1.6$ versions of this fit are also reproduced to show the evolution in the observed BPT diagram. The faint red points show individual \GALFORM galaxies. The larger symbols with errorbars show the medians and 10-90 percentile range for these samples as labelled.}
\label{fig:bpt_z2.3}
\end{figure}

The final comparison is with the Keck Baryonic Structure Survey BPT diagram from \cite{Steidel:2014} at $z \sim 2.3$. 

We use the line ratios given in Tables 1 and 2 from \cite{Steidel:2014}; the latter gives upper limits on the [NII]/H${\alpha}$ ratio whereas the former gives measurements. These objects are plotted as green crosses in Fig.~\ref{fig:bpt_z2.3}. Note some of the objects in the top right corner of the BPT diagram were classified as AGN by \cite{Steidel:2014}. Around half a dozen star-forming galaxies in the \cite{Steidel:2014} sample lie just above the maximum starburst line from \cite{Kewley:2001}. A significant number of emission line galaxies classified as star forming lie above the \cite{Kauffmann:2003} line which separates local AGN from star-forming galaxies. The green-blue dashed line shows the relation from \cite{Faisst:2018} at $z=2.34$, which was calibrated using these observations.
The best fit to the BPT locus of star forming galaxies at $z=2.34$ given by \cite{Steidel:2014} is shown by the solid green line in Fig.~\ref{fig:bpt_z2.3}, and agrees well with the Faisst et~al. curve (which included these observations in its calibration).  

Fig.~\ref{fig:bpt_z2.3} also shows \GALFORM galaxies at $z=2.34$ which have [OIII] and H$_{\beta}$ fluxes greater than $3.5 \times 10^{-18} {\rm erg \,s}^{-1} {\rm cm}^{-2}$ and [NII] and H$_{\alpha}$ fluxes greater than $5.0 \times 10^{-18} {\rm erg \,s}^{-1} {\rm cm}^{-2}$. 
The median [OIII]/H$\beta$ ratio declines more sharply in the model with increasing [NII]/H$\alpha$ than it does for the observations, but this again reflects the difficulty in reproducing the observational selection. There are no observed galaxies with $\log$[OII]/H$\beta < 0$.

\subsection{Evolution of the properties of HII regions} 

 \begin{figure}
    	\includegraphics[clip, trim=0.cm 0cm 0cm 0.0cm,width=0.82\columnwidth]{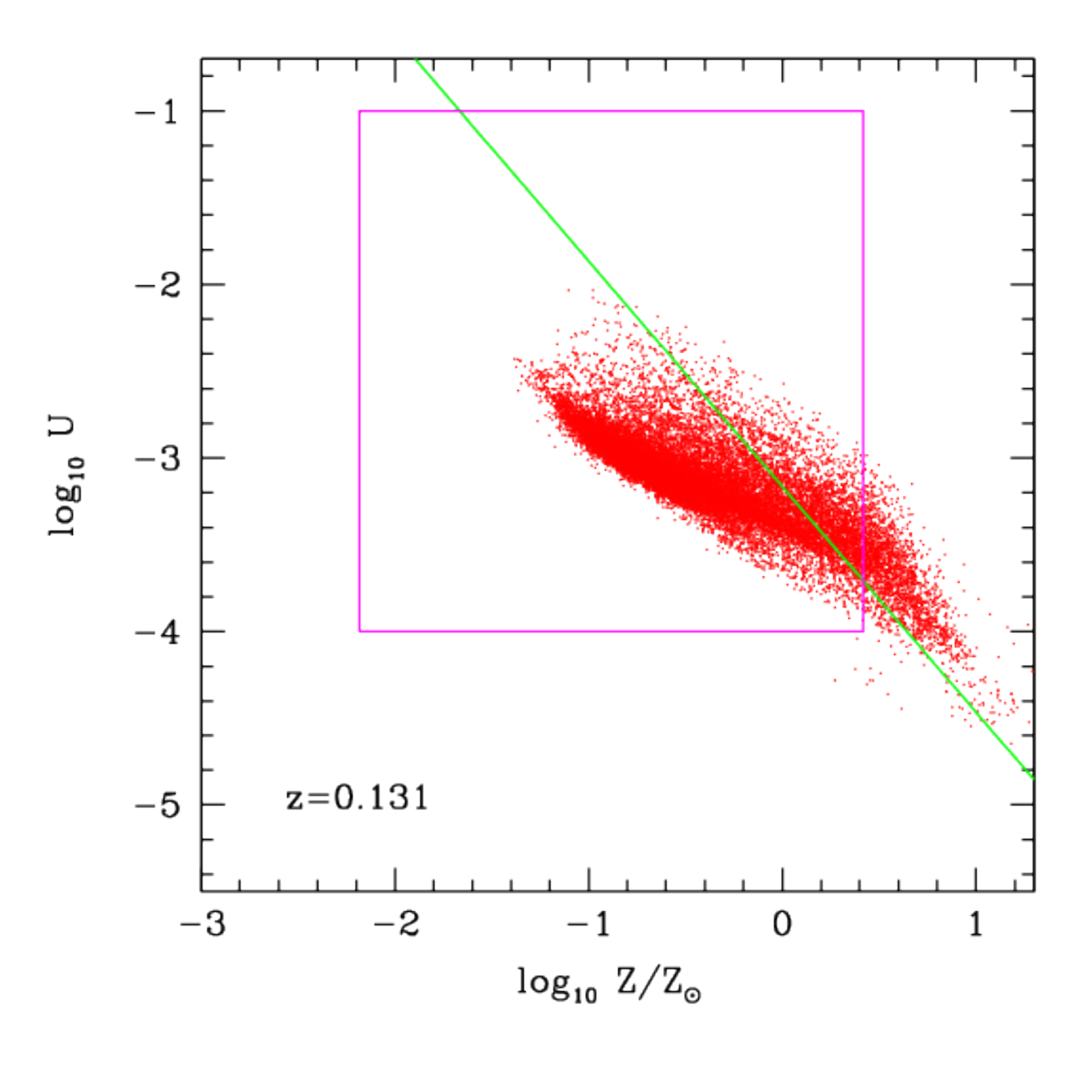}
    	\includegraphics[clip, trim=0.cm 0cm 0cm 0.5cm,width=0.82\columnwidth]{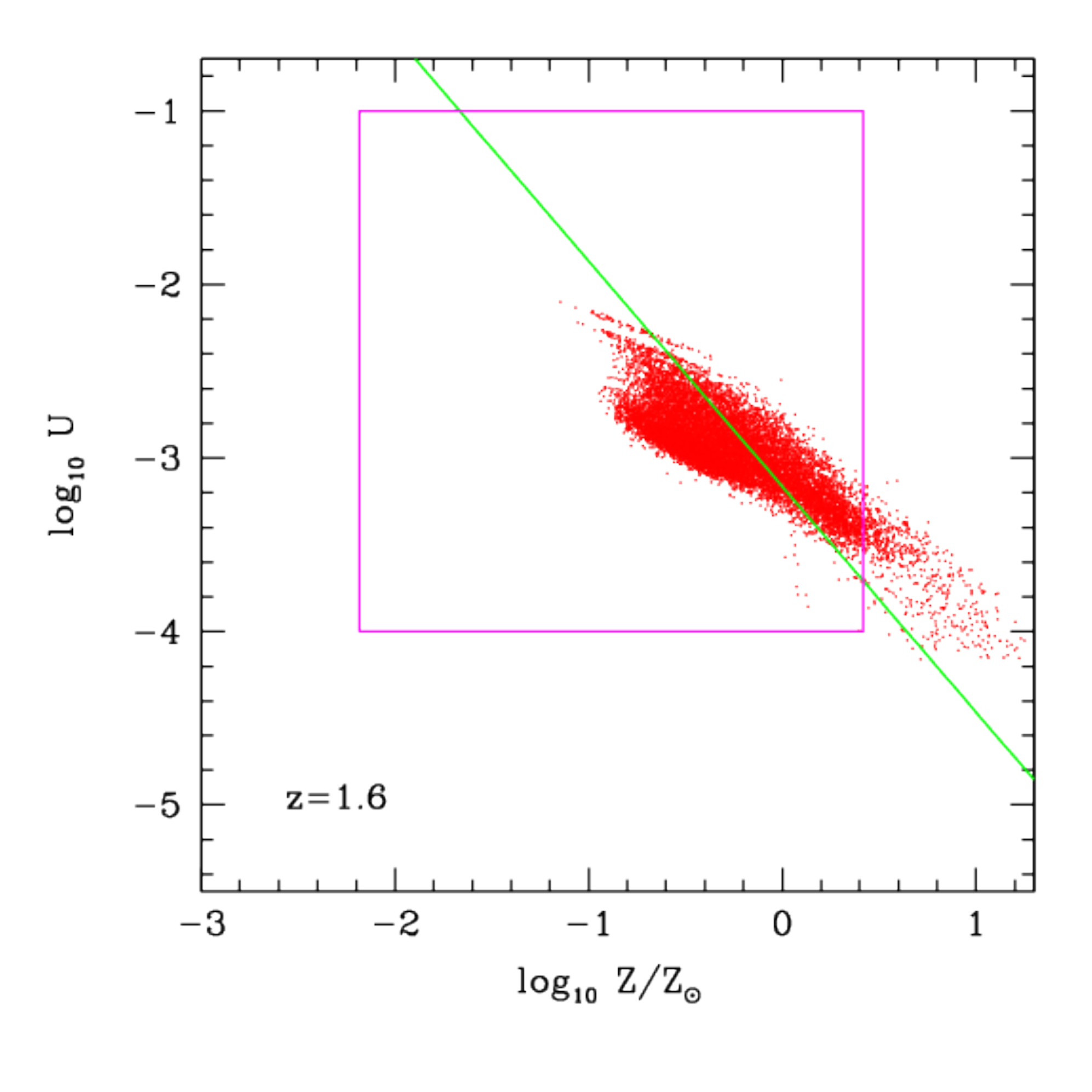}	\includegraphics[clip, trim=0.cm 0cm 0cm 0.5cm,width=0.82\columnwidth]{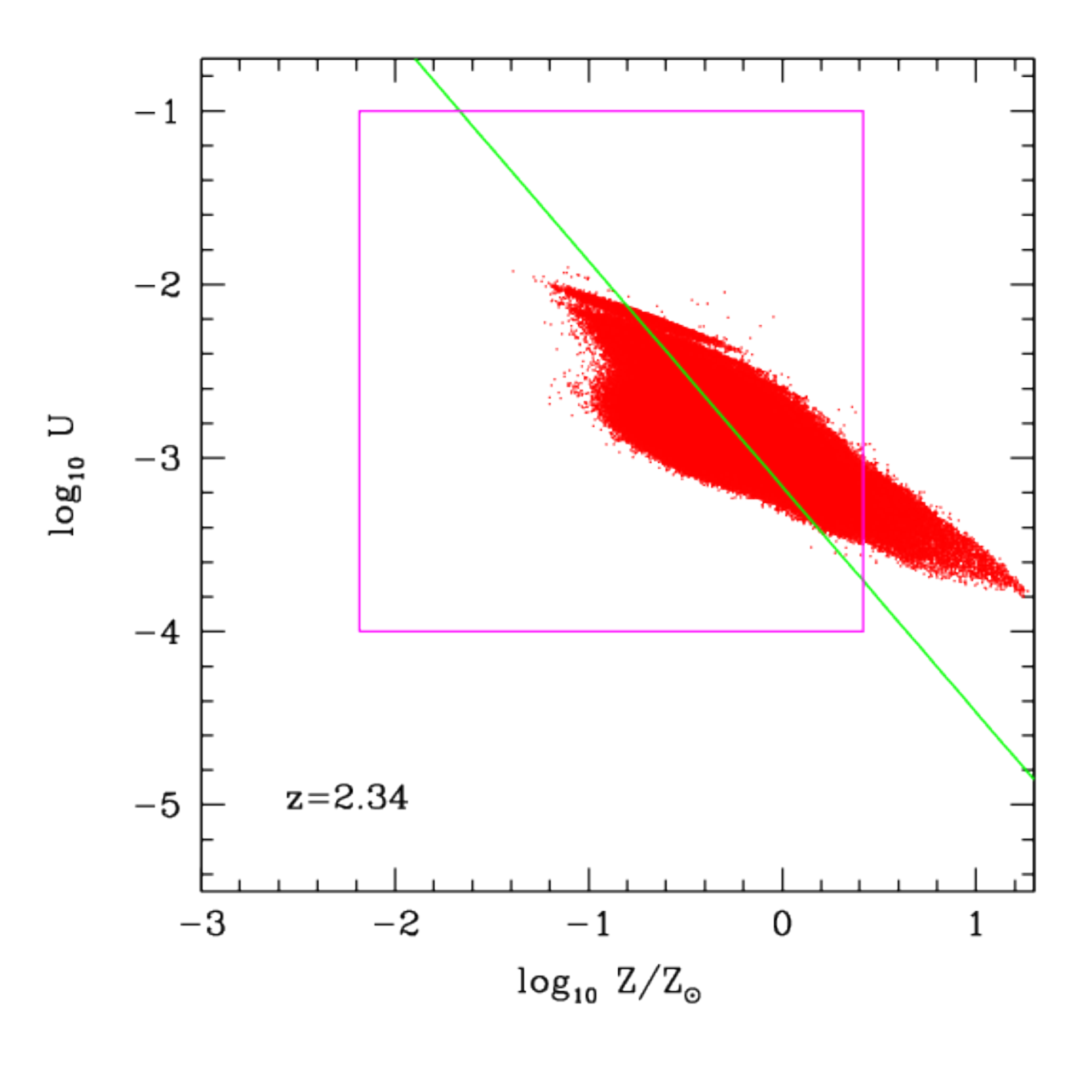}
    \caption{The ionisation parameter, $U$, and metallicity of the star forming gas, $Z$, in solar units, in the \GALFORM  samples (red points) at $z=0.131$ (top), $z=1.6$ (middle) and $z=2.34$ (bottom). The model galaxies have met the various observational selections for each redshift described in the text. The green line shows the ionisation - metallicity model used by Orsi et~al. (2014). The magenta box shows the extent of the parameter space in the  precomputed grid of HII models from Gutkin et~al. (2016).
  }
    \label{fig:uzgrid}
\end{figure}

\begin{figure}
    	\includegraphics[clip, trim=0.75cm 1.5cm 0cm 0.5cm,width=0.99\columnwidth]{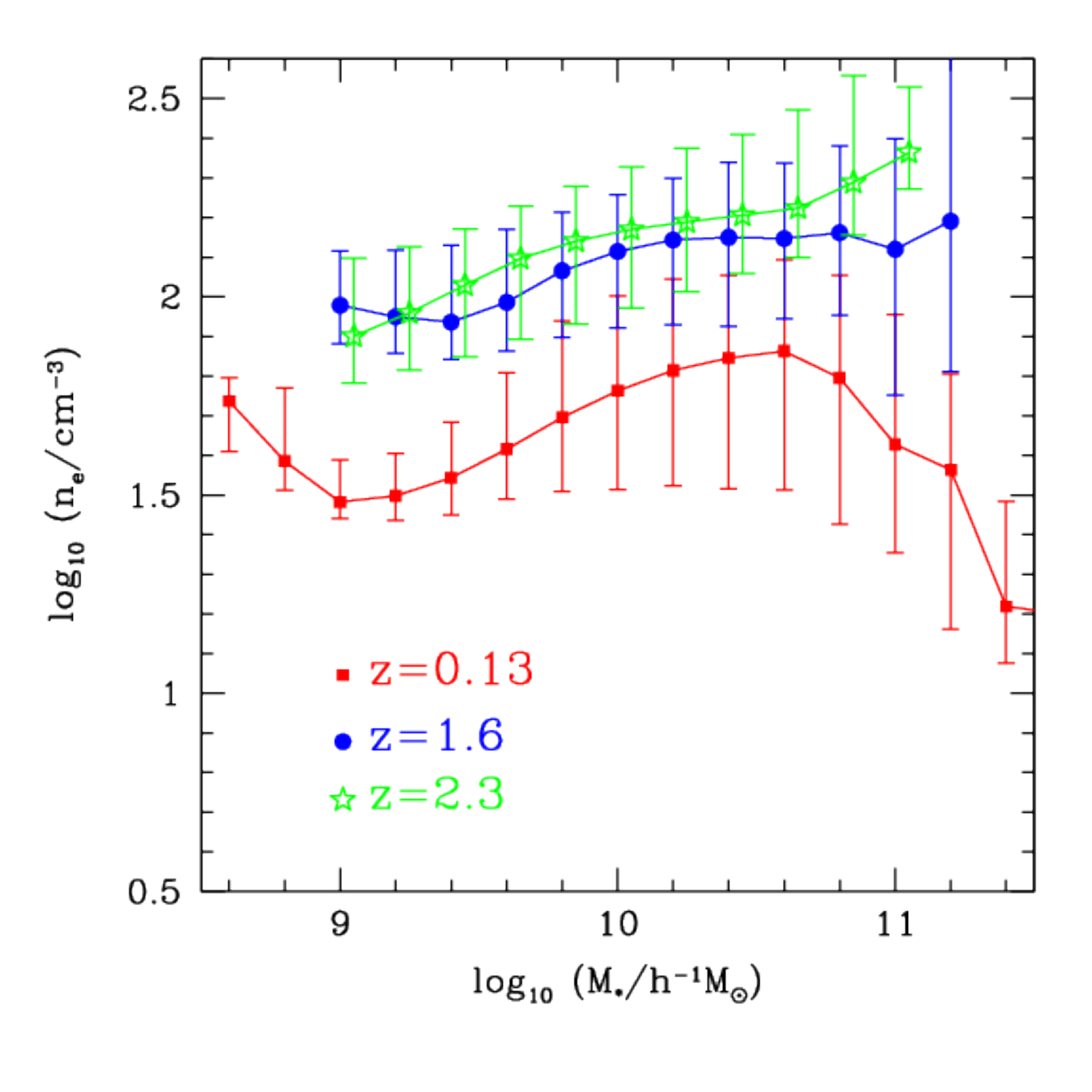}
    	\includegraphics[clip, trim=0.75cm 1.5cm 0cm 0.5cm,width=0.99\columnwidth]{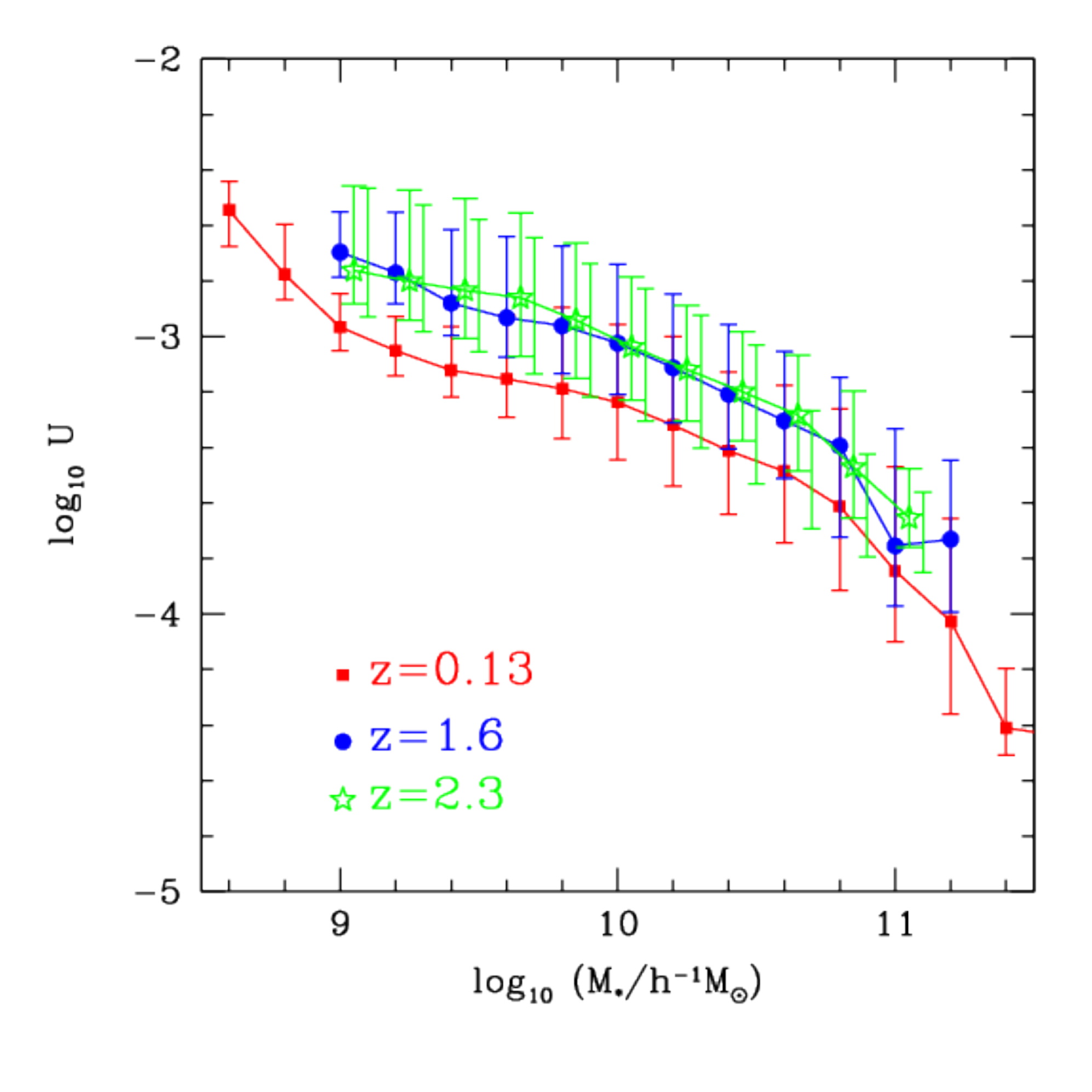}
    \caption{ Top: the evolution of the electron density adopted in the model HII regions, plotted as a function of galaxy stellar mass. The electron density is computed from the galaxy's stellar mass and specific star formation rate using Eqn.~\ref{eq:ne}. Different colours indicate different redshifts as labelled; note that the observational selection applied to the model galaxies varies with redshift. Bottom: the evolution of the ionisation parameter $U$, again plotted against stellar mass. Lines have the same meaning as in the top panel.
  }
    \label{fig:nez}
\end{figure}

\begin{figure}
    	\includegraphics[trim=0.75cm 1.5cm 0cm 0.5cm,clip,width=0.99\columnwidth]{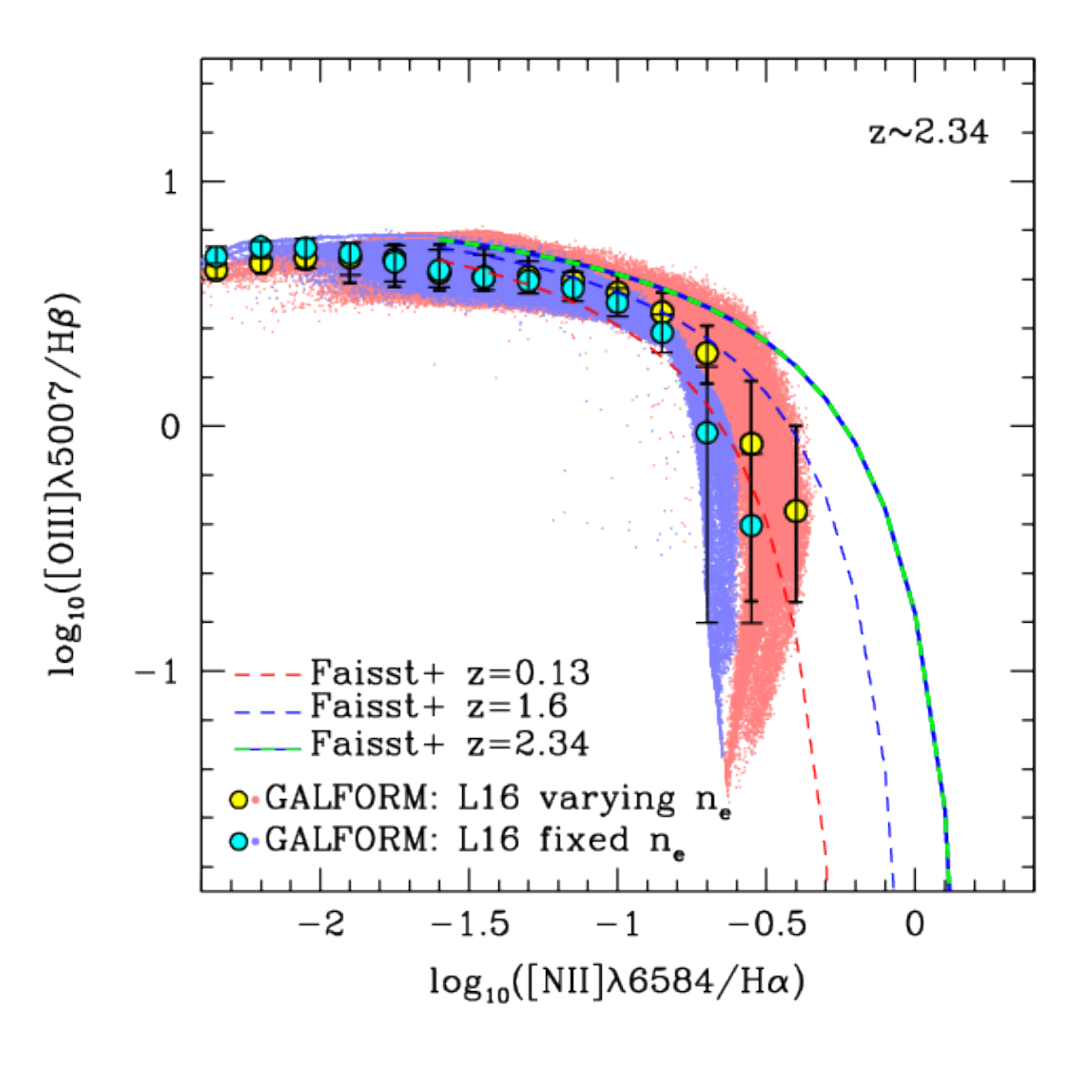}
        \includegraphics[trim=0.75cm 1.5cm 0cm 0.5cm,clip,width=0.99\columnwidth]{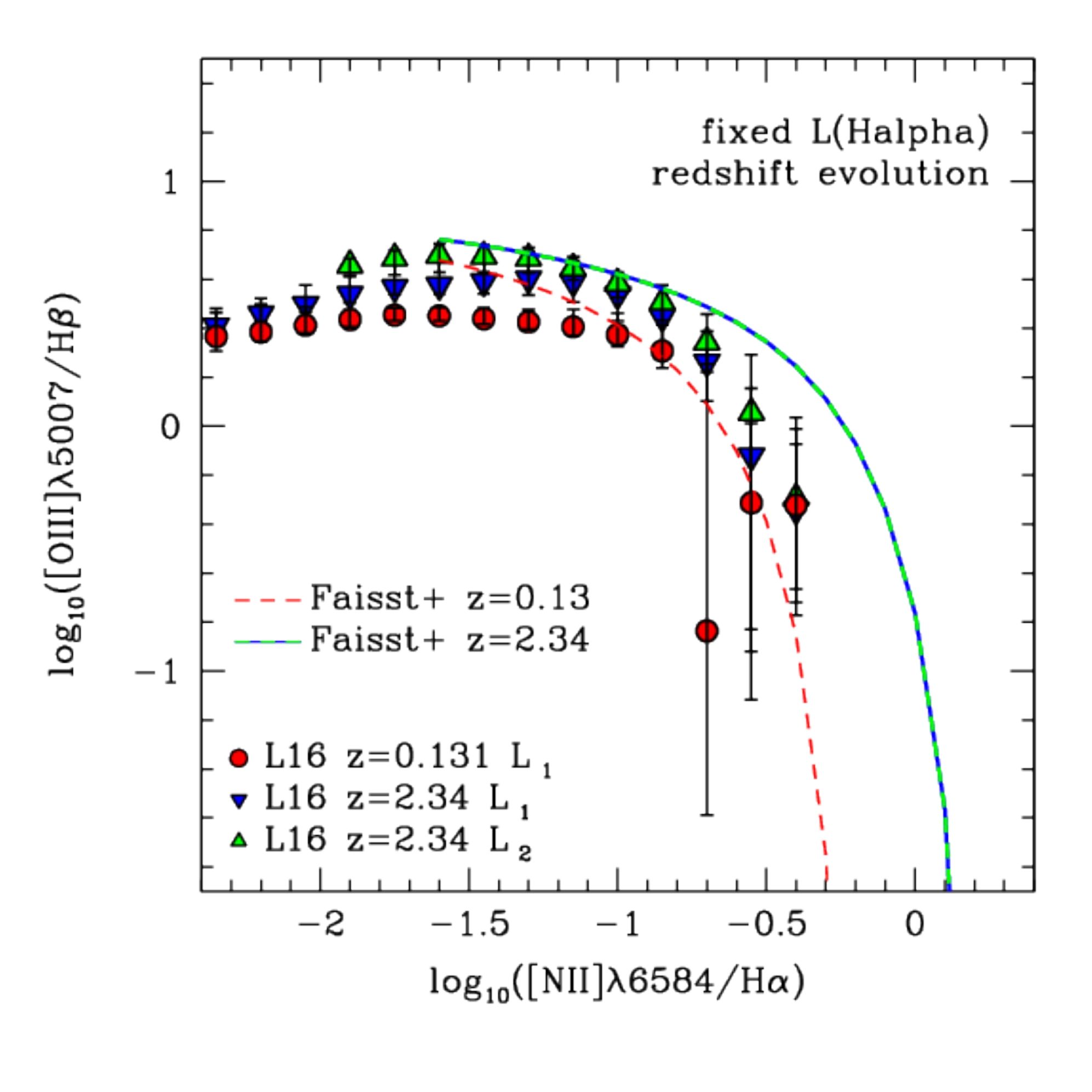}	
    \caption{ Top: the prediction of our fiducial model, in which the gas density is allowed to vary (pink dots show individual galaxies, yellow filled points show median line ratios), compared with a variant model in which the gas/electron density is held fixed at $\log(n_{\rm e}/{\rm cm}^{-3}) = 1.5$. Bottom: comparing the predicted evolution in the BPT diagram for different cuts on H$\alpha$ luminosity (all other observational selections have been removed). Note that in this panel, we allow the gas density to vary, as in our standard model. Two luminosities are considered: $L_{1}$ at both low and high redshift, which is the luminosity cut applied to GAMA galaxies, and $L_{2}$, which reflects the change in the flux limit and luminosity distance for the $z=2.3$ sample.  
  }
    \label{fig:bpt_evol}
\end{figure}

Our new model uses the empirical connection between the properties of individual HII regions and the global properties of galaxies derived by \cite{Kashino:2019b} from observations of local galaxies. \GALFORM \, models the chemical evolution of baryons and predicts the metallicity of the star forming gas, in the disk for quiescent star formation and in the bulge component for bursts of star formation.   
\cite{Kashino:2019b} found that the ionisation parameter $U$ depends on the gas metallicity, the electron or gas density in the HII region and the specific star formation rate. \cite{Kashino:2019b} give a fitting formula  for the electron density in terms of the stellar mass of the galaxy and its specific star formation rate (Eqn.~\ref{eq:ne}; see their Table 2). In this way, the electron density and hence the ionisation parameter can be computed from global galaxy properties that are predicted by \GALFORM. 

The predictions of the L16 model for the ionisation parameter, $U$, and metallicity, $Z$, of HII regions are shown in Fig.~\ref{fig:uzgrid}\footnote{The GP14 model overlaps with these predicted properties, without covering the same range of the parameters, i.e. the L16 HII region properties extend to lower and higher metallicities than those in the GP14 model, particularly at high redshift.}. Each panel shows one of the redshifts at which the model was compared to observations in the preceding subsections. In each case, as set out above, different observational selections have been applied to the model galaxies. The grid of values for which the precomputed HII models are available is shown by the magenta box, which is reproduced in each panel of Fig.~\ref{fig:uzgrid}. Galaxies which lie outside this box are assigned the HII model corresponding to the parameter value at the closest edge of the box.  For comparison, the model proposed by \cite{Orsi:2014} (Eqn.~\ref{eq:qgetemlines}) is shown by the green line (see Appendix B for a discussion of this model). The 
Orsi et al. model adopts a monotonic relation between the ionisation parameter and metallicity, without any scatter. This relation is steeper than the one we recover using the Kashino \& Inoue fits in \GALFORM . 

The evolution of the electron density and ionisation parameter predicted in \GALFORM \, galaxies using Eqns.~\ref{eq:ne} and 3 is shown in Fig.~\ref{fig:nez}. According to Eqn.~\ref{eq:ne}, the electron density in the HII region depends on the stellar mass and specific star formation rate in a model galaxy, although the resulting dependence on stellar mass in Fig.~\ref{fig:nez} is modest. An additional influence over the  apparent evolution in electron density shown in Fig.~\ref{fig:nez} is the different selection criteria applied at each redshift to mimic those used in the observational samples. The electron density changes by factor of $\approx 3$ depending on the stellar mass between $z=0.13$ and $z=1.6$, with only a modest further increase on moving to $z=2.3$. The ionisation parameter declines sharply with increasing stellar mass, as shown by the bottom panel in Fig.~\ref{fig:nez}. As with the density in the HII regions, the ionisation parameter increases with increasing redshift. At a stellar mass of $3 \times 10^{10} h^{-1} M_{\odot}$, the ionisation parameter increases by a factor of almost $2$ between $z=0.13$ and $z=2.3$.

We explore the reasons behind the evolution in the model BPT diagram further in Fig.~\ref{fig:bpt_evol}. In the top panel, we compare the predictions of our fiducial model, in which the gas density is allowed to vary, with a version of the model in which the density is held fixed at $\log (n_{\rm e}/{\rm cm}^{-3}) = 1.5$. This value is approximately the median gas density predicted in the model at $z=0.13$. Allowing the gas density to vary increases the [OIII]/H$\beta$ ratio by roughly a factor of two for high values of [NII]/H$\alpha$, but has no effect at lower values of [NII]/H$\alpha$. In the lower panel of Fig.~\ref{fig:bpt_evol} we remove all selection criteria apart from H$\alpha$ luminosity to compare the lowest and highest redshift samples, and revert to our standard assumption of allowing the gas density in the HII region to vary. Considering objects with the same intrinsic H$\alpha$ luminosity, equivalent to the luminosity cut applied in the GAMA comparison, we see a shift in the predicted BPT diagrams between $z=0.13$ and $z=2.3$ (i.e. comparing the samples labelled $L_{1}$). This evolution for objects with the same H$\alpha$ luminosity is produced by the change in ionisation parameter shown in Fig.~\ref{fig:nez}. The change in the luminosity distance between these redshifts means that to compare at a fixed \textit{flux} rather than luminosity, we should consider objects that are $\approx 1\,000$ times brighter at $z=2.3$ than their $z=0.13$ counterparts. However, the flux selection applied in the Keck sample is $\approx 50$ times fainter than the GAMA flux limit, so it is more relevant to compare with high redshift galaxies whose intrinsic luminosity is 20 times brighter than the low redshift ones. These galaxies are shown by the green triangles in Fig.~\ref{fig:bpt_evol}, which show a further small increase in [OIII]/H$\beta$ line ratios, over that seen on changing the redshift at a fixed luminosity. The evolution in the model BPT diagram is driven by a combination of an increase in the luminosity of galaxies selected with increasing redshift, an increase in their ionisation parameter with redshift (which is apparent even at fixed luminosity) and by allowing the gas density to vary, which alters the shape of the locus of the BPT region. With these changes in the properties of the low and high redshift galaxies and their HII regions, the model is able to reproduce the observed evolution in the position of star-forming galaxies in the BPT diagram to $z\sim1.6$, but not the additional evolution seen to $z=2.3$. We have explored increasing the H$\alpha$ cut further to reproduce the observed evolution to $z=2.3$, but this requires applying a flux cut that is substantially larger than the one applied to the observations.

\section{Summary and conclusions}
We have introduced a new model to predict the emission lines from star forming galaxies, which we have implemented in the \GALFORM\, semi-analytical model of galaxy formation. We use the empirical relations connecting the properties of HII regions with the global properties of galaxies derived from local galaxies by \cite{Kashino:2019b} and combine this with the pre-computed grid of HII region models from \cite{Gutkin:2016}. Our new model reproduces the local BPT relation for star forming galaxies. We recover the evolution seen in the locus of star-forming galaxies in the BPT diagram to $z=1.6$, but not the additional evolution seen to $z=2.3$ (e.g. \citealt{Faisst:2018}). The model predicts that the high redshift star forming galaxies probed in emission line samples have gas densities and ionisation parameters that are $\approx 2-3$ times higher than in their local counterparts.

There are three directions in which the new model introduced here could be improved in future work.  

The first is to fully couple the HII region model with the stellar population synthesis model used in the galaxy formation calculation. At present, the connection is made through the production rate of Lyman continuum photons. Different stellar population models can be (and are) used in calculating the SED of the stellar population in a galaxy and in the photoionization modelling of the nebular emission. The composite spectra of the galaxies could differ in the ultra-violet due to the treatment of different stages of stellar evolution in the respective stellar population synthesis models (see for example the comparison of stellar population synthesis models in \citealt{Violeta:2014}). 
Also, if a top-heavy IMF is assumed in starbursts, as in the L16 model, this will result in a harder spectrum for the ionizing photons, {\it in addition} to a higher production rate of ionizing photons for a given star formation rate.
Finally, by using a pre-computed HII region grid that is consistent with the properties of the \GALFORM galaxies, the HII model grid could readily be extended to cover the full range of gas metallicities predicted by the galaxy formation model. The grids currently used do not extend to the highest metallicities produced in the model; in this case we use the HII region model at the upper limit of the available metallicity grid, rather than trying to extrapolate the HII model grid output to a higher metallicity.  

The second improvement concerns the connection between the properties of the HII regions and the global interstellar medium in a galaxy, which is based on the analysis of local galaxies by \cite{Kashino:2019b}. Our model implicitly assumes that the HII regions in high redshift galaxies are extreme counterparts of local star forming galaxies, perhaps with higher specific star formation rates, electron/gas densities and ionisation parameters than a `typical' local star forming galaxy. This interpretation is supported by some observational studies. \cite{Kaasinen:2017} found that the electron density in a sample of galaxies at $z \sim 1.5$ is five times higher than that of local galaxies. However, star formation rate is a confounding variable and when samples are compared at the same star formation rate they are found to have the same electron density (see also \citealt{Liu:2008,Kaasinen:2018}).
Several studies have argued that simply selecting the extreme tail of the local star-forming galaxy population is insufficient to explain the evolution of the BPT diagram and that there are other differences that drive this behaviour \citep{Steidel:2014, Bian:2020}.  It would be instructive to repeat the analysis carried out by  \cite{Kashino:2019b} for the samples of high redshift galaxies used to establish the evolution in the BPT diagram, to see if the connection between HII region properties and galaxy properties evolves in a way not described by the $z=0$ relations.  

The third way in which the model could be extended is by implementing different stellar population synthesis models which produce harder spectra for star forming galaxies, as suggested by \cite{Steidel:2014}. One possibility is the models that take into account that most stars are part of binary systems and that the presence of a binary can prolong the emission of ionising photons  
\citep{Stanway:2014,Xiao:2018}.

The model presented here is a first step to producing a more physical model of emission lines in star forming galaxies. By coupling the galaxy formation model to a HII region model and linking the predicted galaxy properties to those of the HII regions, we will be able to use upcoming observations, for example from DESI\footnote{https://www.desi.lbl.gov/}, Euclid\footnote{https://www.cosmos.esa.int/web/euclid}, the Webb Telescope\footnote{https://www.nasa.gov/mission\_pages/webb/main/index.html} and the MOONS spectrograph\footnote{https://vltmoons.org/}, to probe the model predictions for the state of the ISM in galaxies at the peak epoch of cosmic star formation and beyond.

\section*{Acknowledgements}
CMB acknowledges conversations with P. Norberg and N. Padilla, and A. Orsi for providing a copy of {\tt GET\_EMLINES}. The authors acknowledge the contributions of Jack Humphries and Michele Fumagalli to a precursor to this project. 
This work was supported by the Science and Technology Facilities Council [ST/P000244/1]. VGP is supported by the Atracci\'{o}n de Talento Contract no. 2019-T1/TIC-12702 granted by the Comunidad de Madrid in Spain. This work used the DiRAC@Durham facility managed by the Institute for Computational Cosmology on behalf of the STFC DiRAC HPC Facility (www.dirac.ac.uk). The equipment was funded by BEIS capital funding via STFC capital grants ST/K00042X/1, ST/P002293/1, ST/R002371/1 and ST/S002502/1, Durham University and STFC operations grant ST/R000832/1. DiRAC is part of the National e-Infrastructure. This research has made use of NASA’s Astrophysics Data System.
This paper made use of data from GAMA DR3. GAMA is a joint European-Australasian project based around a spectroscopic campaign using the Anglo-Australian Telescope. The GAMA input catalogue is based on data taken from the Sloan Digital Sky Survey and the UKIRT Infrared Deep Sky Survey. Complementary imaging of the GAMA regions is being obtained by a number of independent survey programmes including GALEX MIS, VST KiDS, VISTA VIKING, WISE, Herschel-ATLAS, GMRT and ASKAP providing UV to radio coverage. GAMA is funded by the STFC (UK), the ARC (Australia), the AAO, and the participating institutions. The GAMA website is http://www.gama-survey.org/ .

\section*{Data availability}
The model output presented in this paper is available from the corresponding author on request.


\bibliographystyle{mnras}
\bibliography{emission_lines} 

\appendix
\section{The original GALFORM emission line model} 

\begin{figure}
\includegraphics[clip, trim=0.cm 0cm 0cm 0.0cm,width=0.98\columnwidth]{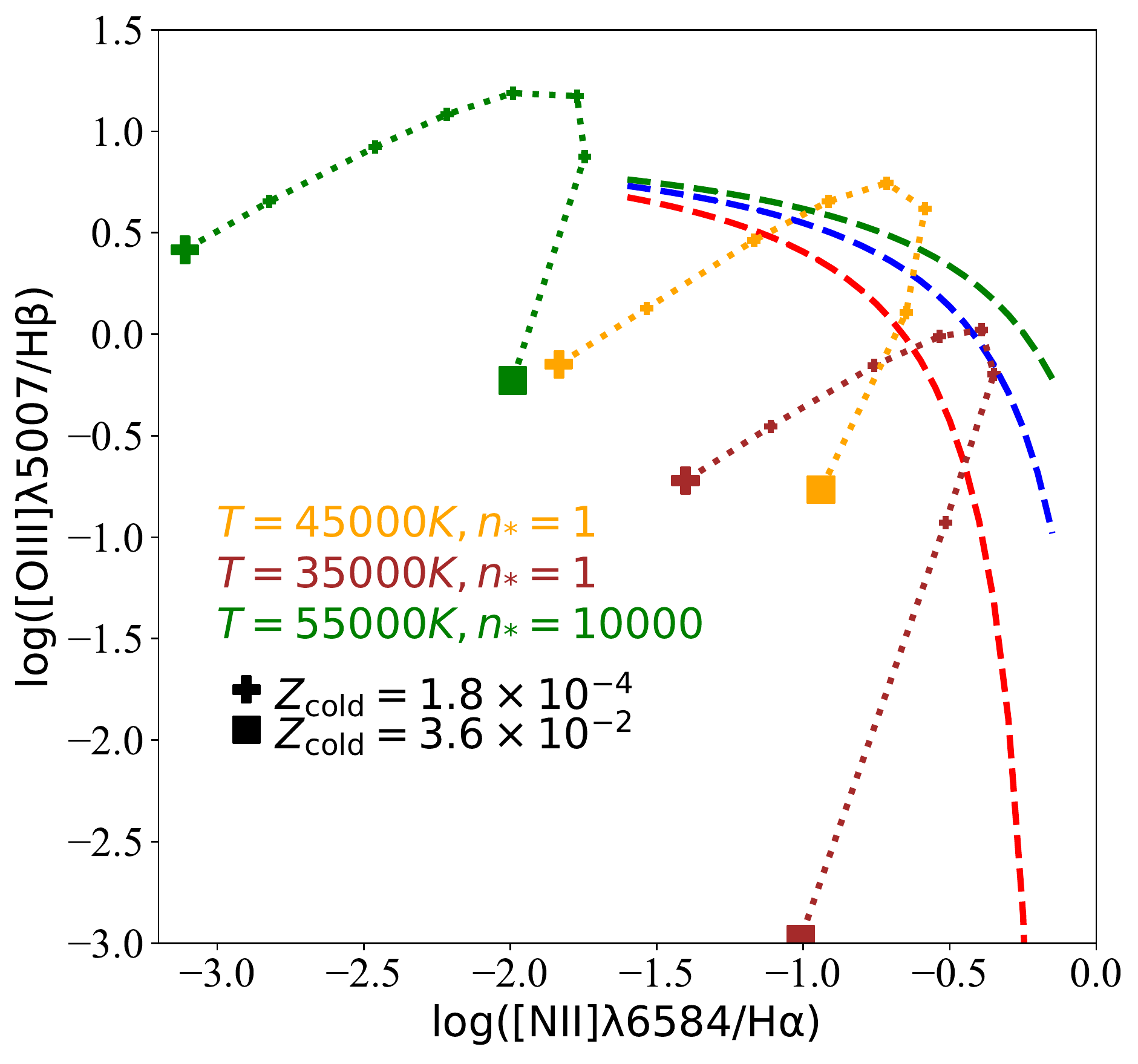}
    \caption{The BPT diagram showing selected HII model predictions from Stasi\'{n}ska (1990). 
    The orange dotted curve shows the default model used in \GALFORM  which corresponds to one ionising star per HII region with a temperature of $45\,000$K.  The brown dotted curve shows a variant model in which the ionising star has a lower effective temperature. The green dotted curve shows a more extreme case in which there are 10\,000 ionising stars, each with a higher effective temperature. In all cases the gas density is $n_{\rm g} = 10 \, {\rm cm}^{-3}$. The symbols indicate the sequence of metallicity along the curves, moving from the lowest metallicity (plus sign) to highest value (square), as labelled. The dashed lines show fits to observational measurements at $z=0.13$ (red), $z=1.6$ (blue) and $z=2.34$ (green) using the parametrization from Faisst et~al. (2018). 
    These curves are plotted over the range of the BPT diagram covered by  observational data.}
    \label{fig:bpt_stas}
\end{figure}

Several papers have used the original emission line model implemented in \GALFORM  to make  predictions for Ly$\alpha$ \citep{LeDelliou:2005, LeDelliou:2006, Orsi:2008}, H$\alpha$ \citep{Orsi:2010,Lagos:2014} and OII emitters  \citep{Comparat:2015,Violeta:2018,Violeta:2020}.

The original emission line model takes as input (i) the metallicity of the star forming gas as predicted by \GALFORM, $Z_{\rm cold}$, for star formation taking place either in quiescent disks or bursts, and (ii) the number of Lyman continuum photons emitted per second, $N_{\rm Ly}$, computed from the stellar population synthesis model used in the model. These inputs were then combined with the grid of HII region models from \cite{Stas:1990} to compute the luminosities and equivalent widths for a small number of emission lines. 

The \cite{Stas:1990} models span a range of parameters for the HII regions, including the density of Hydrogen, the number of ionising stars and their effective temperature. The default model selected for use in \GALFORM is for 1 ionising star with an effective temperature of $45 \, 000 \,{\rm K}$ and a Hydrogen density of $10 {\rm cm}^{-3}$. The model results are used in the form of a table with values for the conversion factor $L_{\rm line}/N_{\rm Ly}$ for different cold gas metallicities. The metallicity values range from $1/100^{\rm th}$ to $2 \times$ solar. The emission line calculation is embedded in \GALFORM; however, by outputting the values of $N_{\rm Ly}$ for quiescent star formation in the galactic disk and for any ongoing star burst, along with the cold gas metallicity for these cases, it is possible to compute the emission line luminosities in post-processing, without having to rerun the galaxy formation model. 

The \cite{Stas:1990} model predictions are shown in the BPT diagram in Fig.~\ref{fig:bpt_stas}. This plot shows the default HII model used in \GALFORM (orange curve) and two variants that are sometimes considered, which correspond to different ionising conditions inside the HII region. The \GALFORM model output (not shown) lies along these curves, depending on the distribution of the metallicity of the cold gas. A small number of galaxies show small displacements away from the HII model curve. These galaxies have an ongoing burst {\it and} quiescent star formation in a newly formed disk; the metallicity of the cold gas in these two components can be different, hence the resulting emission line ratios for such a galaxy correspond to the weighted average of two points from this model curve.  

In this model, with a single HII region, there can be no evolution in the locus  of star forming galaxies in the BPT plane with redshift, merely a redistribution of points along the model curve if the metallicity of the cold gas evolves with redshift. Note also that the shape of the HII model curve does not match the locus observed for star forming galaxies. In the model, local galaxies with cold gas metallicities that placed around the apex of the HII model curve would be in danger of being misclassified as AGNs according to the criteria usually used.

\section{GET\_EMLINES}

\begin{figure}
    	\includegraphics[clip, trim=0.cm 0cm 0cm 0.0cm,width=0.98\columnwidth]{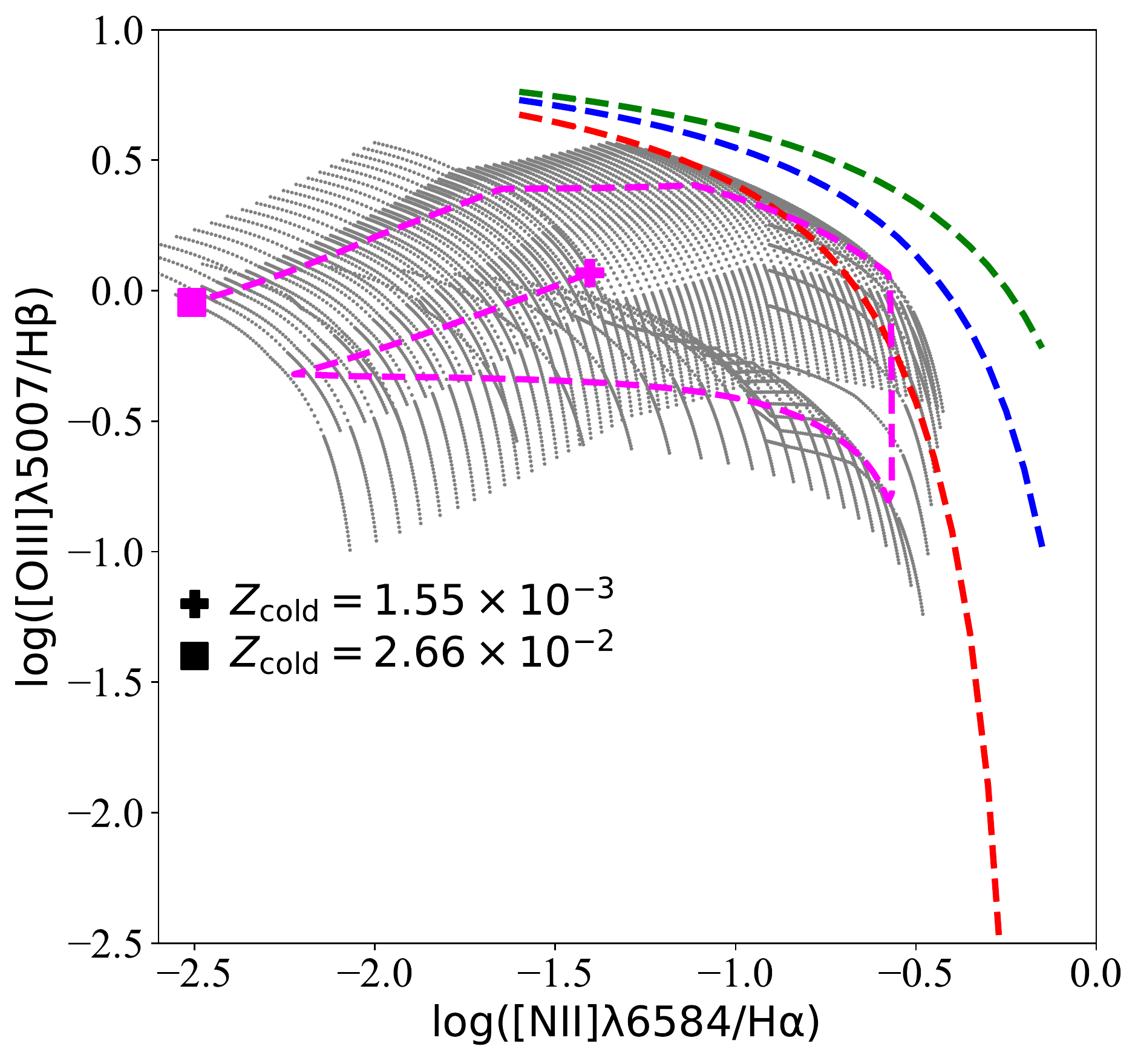}
    \caption{The BPT diagram showing the favoured model from \getemlines (magenta dashed curve). The symbols again indicate the sequence of metallicity along the curves, moving from the lowest (plus sign) to highest value (square), as labelled. The range of metallicities shown is determined by the range of ionisation parameters, $q$, for which the grid of HII models from Levesque et~al. (2010) is provided. The grey dots show the Levesque et~al. grid used by the \getemlines model, varying the gas metallicity, $Z_{\rm cold}$, and the ionisation parameter, $q$, but keeping the age of the stellar cluster and the density of the ionised gas fixed (see text)}. The other dashed lines show fits to observational measurements at $z=0$ (red), $z=1.6$ (blue) and $z=2.3$ (green) taken from Faisst et~al. (2018). Note the change in the axis limits compared with Fig.~\ref{fig:bpt_stas}.
    \label{fig:bpt_getemlines}
\end{figure}

\cite{Orsi:2014} introduced the \getemlines
\footnote{https://github.com/aaorsi/get\_emlines} model of emission lines and implemented this into the {\tt SAG} semi-analytical galaxy formation model of \cite{Gargiulo:2015}. This model has since been used with  other semi-analytic models ({\tt SAG, SAGE, Galacticus}) by \cite{Favole:2019ouk} to make predictions for OII emitters and  by \cite{Izq:2019} with {\tt L-GALAXIES} to make mock catalogues. 

The \getemlines model uses a grid of HII models from \cite{Levesque:2010}. These models were calculated by feeding simple stellar populations from the Starburst99 code of \cite{Leitherer:1999} (assuming a \citealt{Salpeter:1955} IMF) into the {\tt MAPPINGS-III} photoionsation code of \cite{Groves:2004}.

The model grid of \cite{Levesque:2010} covers four parameters:
(i) the age of the stellar cluster that provides the ionising radiation, (ii) the density of the ionised gas, (iii) the metallicity of the cold gas, and (iv) the ionisation parameter, $q$, which is simply  related to the dimensionless ionisation parameter, $U$, used in the main paper by $q=U c$. 

\cite{Orsi:2014} fixed the first two of these parameters by choosing to adopt a stellar cluster of zero age and an ionised gas density of $10 \,{\rm cm}^{-3}$. The metallicity of the star-forming gas is predicted by the galaxy formation model. Orsi et~al. choose to model the ionisation parameter, $q$, as a function of the metallicity of the cold gas as 
\begin{equation}
q = q_{0} \left( \frac{Z_{\rm cold}}{Z_{0}}\right)^{-\gamma},  \label{eq:qgetemlines}
\end{equation}
where $q_{0}$ and $\gamma$ are parameters; $q_{0}$ is the ionisation parameter at the pivot metallicity $Z_{0}$ which is set to be $Z_{0}=0.012$. Orsi et~al. advocate values for these parameters of $q_{0} = 2.8 \times 10^{7} {\rm cm \, s}^{-1}$ and $\gamma = 1.3$. Orsi et~al. motivated this scaling of the ionisation parameter with metallicity by invoking observational determinations which rely on measuring the ratios of lines from different ionisation states of the same element. These studies suggest that low metallicity galaxies have higher ionisation parameters 
(e.g. \citealt{Groves:2010,Shim:2013,Kashino:2019b}). Orsi et~al. then chose the values of the parameters in Eqn.~\ref{eq:qgetemlines} to match the locus of the observed BPT diagram for star forming galaxies. However, they also showed that their model results for the BPT diagram are fairly insensitive to the precise values of the parameters.
 We note that the {\tt GET\_EMLINES} favoured scaling of $U \propto Z^{-1.3}$, is similar to the apparent correlation found by Kashino \& Inoue, when other quantities are held fixed, but stronger than the scaling with metallicity when the other parameters used by Kashino \& Inoue are also allowed to vary.

By parametrizing the ionising parameter as a function of the metallicity of the cold gas, this reduces the two-dimensional grid of the \cite{Levesque:2010} models to a one-dimensional curve in the BPT diagram. Fig.~\ref{fig:bpt_getemlines} shows the locus of the BPT diagram predicted by \getemlines. There is some overlap between the \getemlines predictions and the observed locus of star-forming galaxies at $z=0$. This model cannot reproduce the evolution in the BPT diagram observed for star forming galaxies, without varying one of the other parameters in the Levesque et~al. grids, such the ionised gas density, with redshift.   
We note that the line ratios predicted by the Levesque et~al. grid do not increase monotonically with $U$ or $Z$. The range of metallicities that can be input into this model is limited by the $q$ range of the Levesque et~al. grid, which is somewhat narrower than the original metallicity range of the grid; this of course depends on the values adopted in Eqn.~\ref{eq:qgetemlines}.

\getemlines takes as input the star formation rate of a galaxy and the metallicity of the star-forming gas; it is straightforward to produce a file suitable for use with \GALFORM by extracting $N_{\rm Ly}$, given the solar neighbourhood stellar initial mass function assumed in \getemlines. To reproduce the behaviour of the model in the BPT diagram accurately, many more metallicity bins ($\approx 50$) are required compared with the case of the \cite{Stas:1990} models. Note that there is some scatter in the {\tt SAG} semi-analytic model predictions presented in \cite{Orsi:2014}. This is not possible for galaxies with a single reservoir of star-forming gas, but is due to objects that are also undergoing a central starburst consuming gas of a different metallicity to that undergoing quiescent star formation in the disk.

Comparing Fig.~\ref{fig:bpt_stas} with Fig.~\ref{fig:bpt_getemlines} we note that the high metallicity point has moved significantly along both axes of the BPT diagram. 

\section{Changing the HII region model} 
\begin{figure}
    	\includegraphics[clip, trim=0.cm 0cm 0cm 0.0cm,width=0.98\columnwidth]{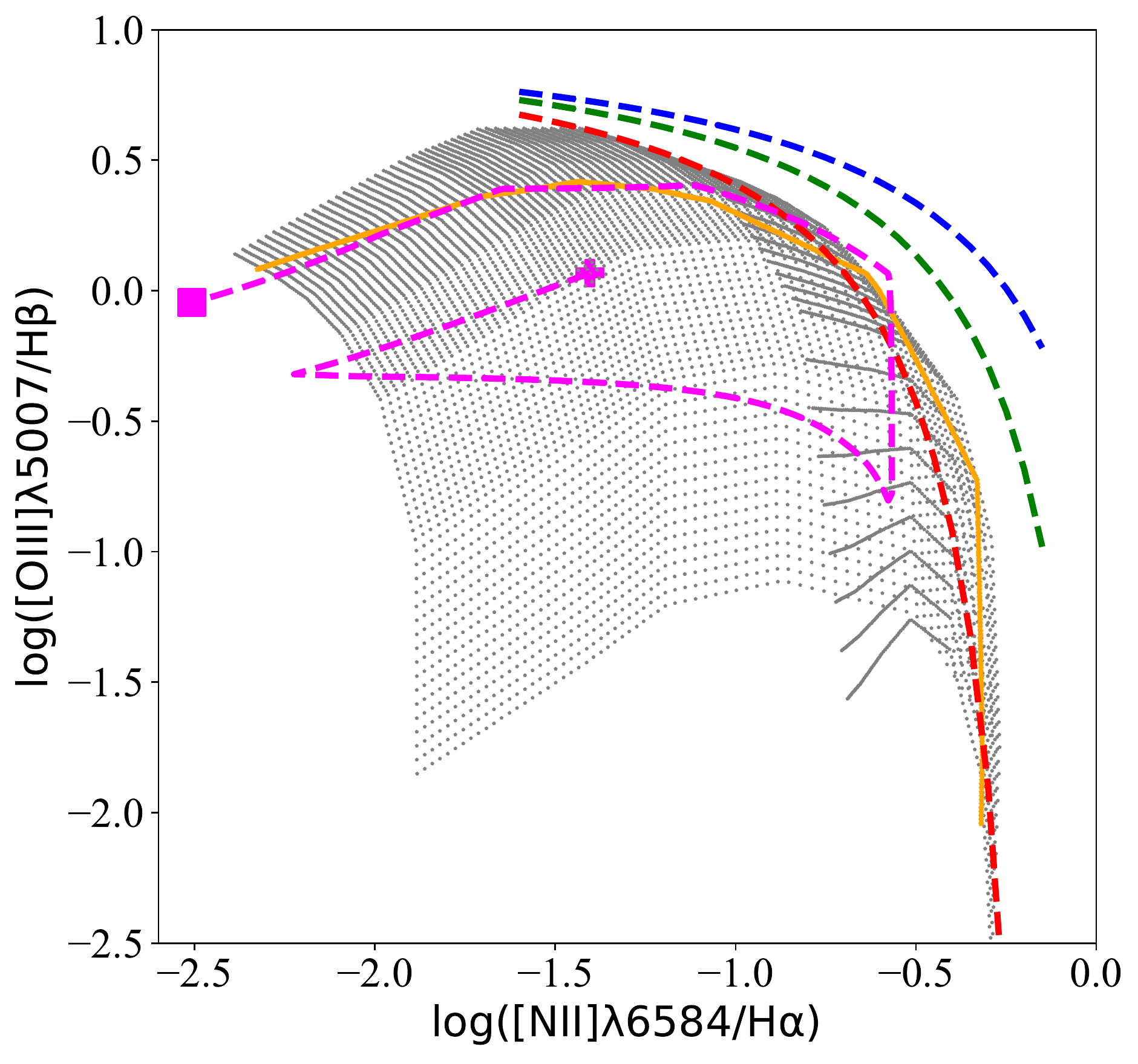}
    \caption{
    The grey dots show the precomputed CLOUDY - FSPS grid from Byler et~al. (2017) in the BPT diagram plane, varying the ionisation parameter, $U$, and gas metallicity $Z$ (for the sense of change of these parameters, see the arrows in Fig.~\ref{fig:bpt_gutkin}). The dashed magenta curve shows the \getemlines model, reproduced from Fig.~\ref{fig:bpt_getemlines}. The solid gold line shows the BPT diagram obtained from the CLOUDY-FSPS model grid when adopting the model for the ionisation parameter advocated by Orsi et~al. (2014). This turns the 2D grid in ionisation parameter and metallicity into a 1D sequence of metallicity. The other dashed lines show fits to observational measurements at $z=0.13$ (red), $z=1.6$ (blue) and $z=2.3$ (green) taken from Faisst et~al. (2018).}
    \label{fig:bpt_fsps_getemlines}
\end{figure}
We now address how the choice of HII region model can affect the predictions for emission line ratios (see also \citealt{Dagostino:2019}). 

First we compare the {\tt MAPPINGS-III} HII model grid with a grid from \cite{Byler:2017}.
\cite{Byler:2017} have combined the Flexible Stellar Population Synthesis ({\tt FSPS}) code of \cite{Conroy:2009,Conroy:2010} with the photoionisation code {\tt CLOUDY} \citep{Ferland:2013} and have provided Python code, {\tt cloudyfsps},  to generate grids of HII models. A pre-computed set of HII model tables is released with FSPS for a solar neighbourhood stellar initial mass function. Although Byler et~al. advise rerunning their calculation for different choices of IMF, we use the original tables (their pre-computed grid assumes a \cite{Kroupa:2001} IMF). 

The Byler et~al. HII models are available for different particle densities and cluster ages, and are parametrized on a grid of ionisation parameter $U$ and cold gas metallicity.

As a simple exercise to let us compare the {\tt CLOUDY FSPS} HII models with those produced using {\tt MAPPINGS-III}, we use the model for $q$ proposed by \cite{Orsi:2014}, as given by Eqn.~\ref{eq:qgetemlines}. We can compare the locus of the {\tt GET\_EMLINES} model on the BPT diagram obtained using the {\tt MAPPINGS-III} models with that obtained using {\tt CLOUDY FSPS}.
The result is shown in Fig.~\ref{fig:bpt_fsps_getemlines}. In this plot, the grid shows the 2D-model of line ratios from the {\tt CLOUDY FSPS} HII model. The plot shows that there is less degeneracy in the line ratios predicted for different combinations of $U,Z$ than is the case with the {\tt MAPPINGS-III grid}. The magenta curve is reproduced from Fig.~\ref{fig:bpt_getemlines} and shows the {\tt GET\_EMLINES} model curve using the Levesque et~al. {\tt MAPPINGS-III} grid. The gold curve in Fig.~\ref{fig:bpt_fsps_getemlines} shows the same model, except using line ratios derived from the {\tt cloudyfsps} grid from Byler et~al. The gold and magenta curves are remarkably similar until low metallicities (i.e. for metallicities below $\approx 5 \times 10^{-3}$ in absolute units). 

We close by noting the {\tt MAPPINGS-III} (Fig.~\ref{fig:bpt_getemlines}) and {\tt CLOUDY-FSPS} (Fig.~\ref{fig:bpt_fsps_getemlines}) reach similar values for the OIII/H$\beta$ ratio, but do not reach the values predicted by the \cite{Gutkin:2016} grid used in the main paper. In particular, the {\tt MAPPINGS-III} and {\tt CLOUDY FSPS} grids do not overlap with the high redshift locii of star forming galaxies on the BPT diagram. We have not been able to isolate the reason for this difference, which, as far as we know has not been discussed in the literature. \cite{Violeta:2014} do show that different stellar population synthesis models predict different spectra in the ultra-violet for the same amount of star formation, which could be part of the reason for this difference. 
 
\section{Other line ratios are available} 
\begin{figure}
    	\includegraphics[clip, trim=0.cm 0cm 0cm 0.0cm,width=0.98\columnwidth]{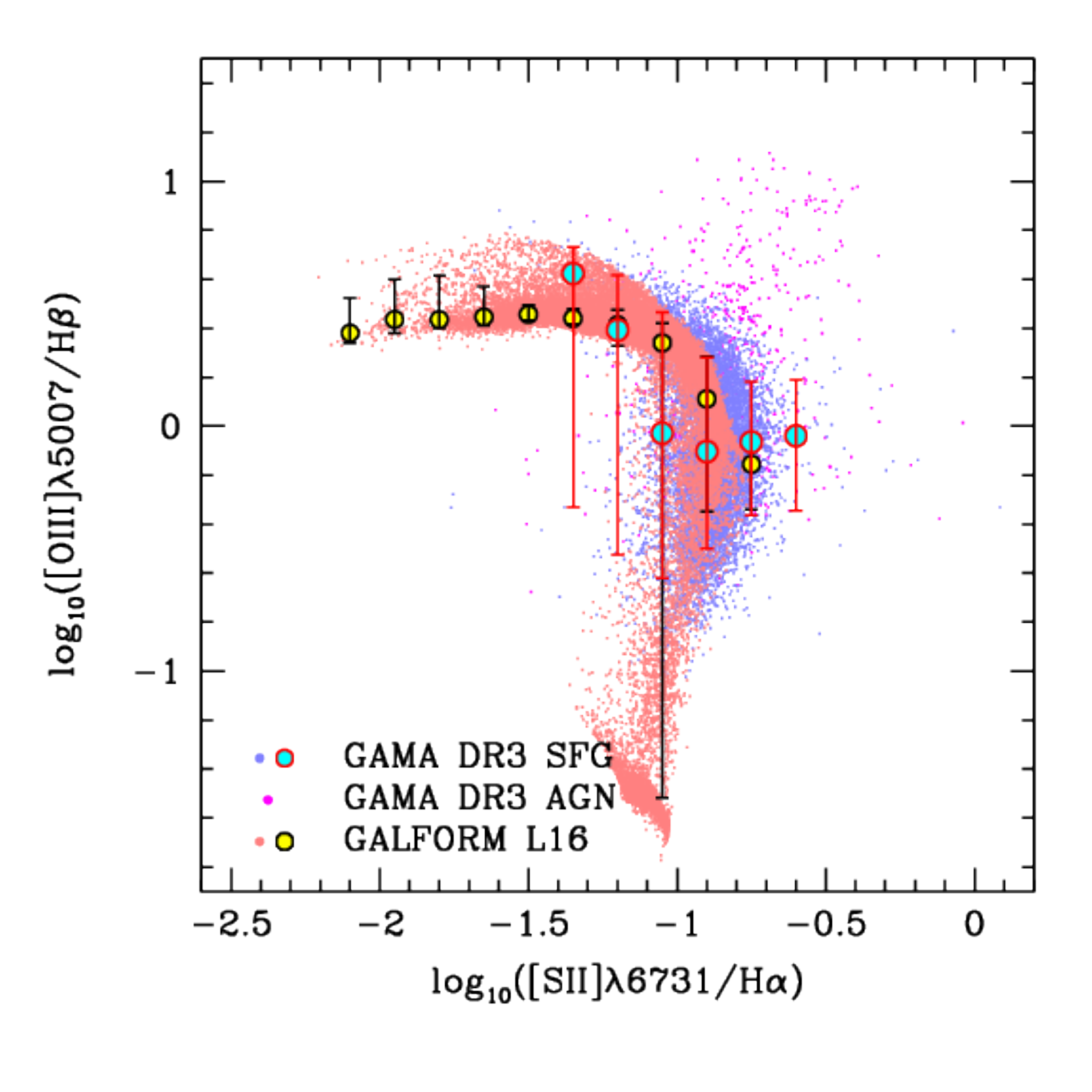}
        	\includegraphics[clip, trim=0.cm 0cm 0cm 0.0cm,width=0.98\columnwidth]{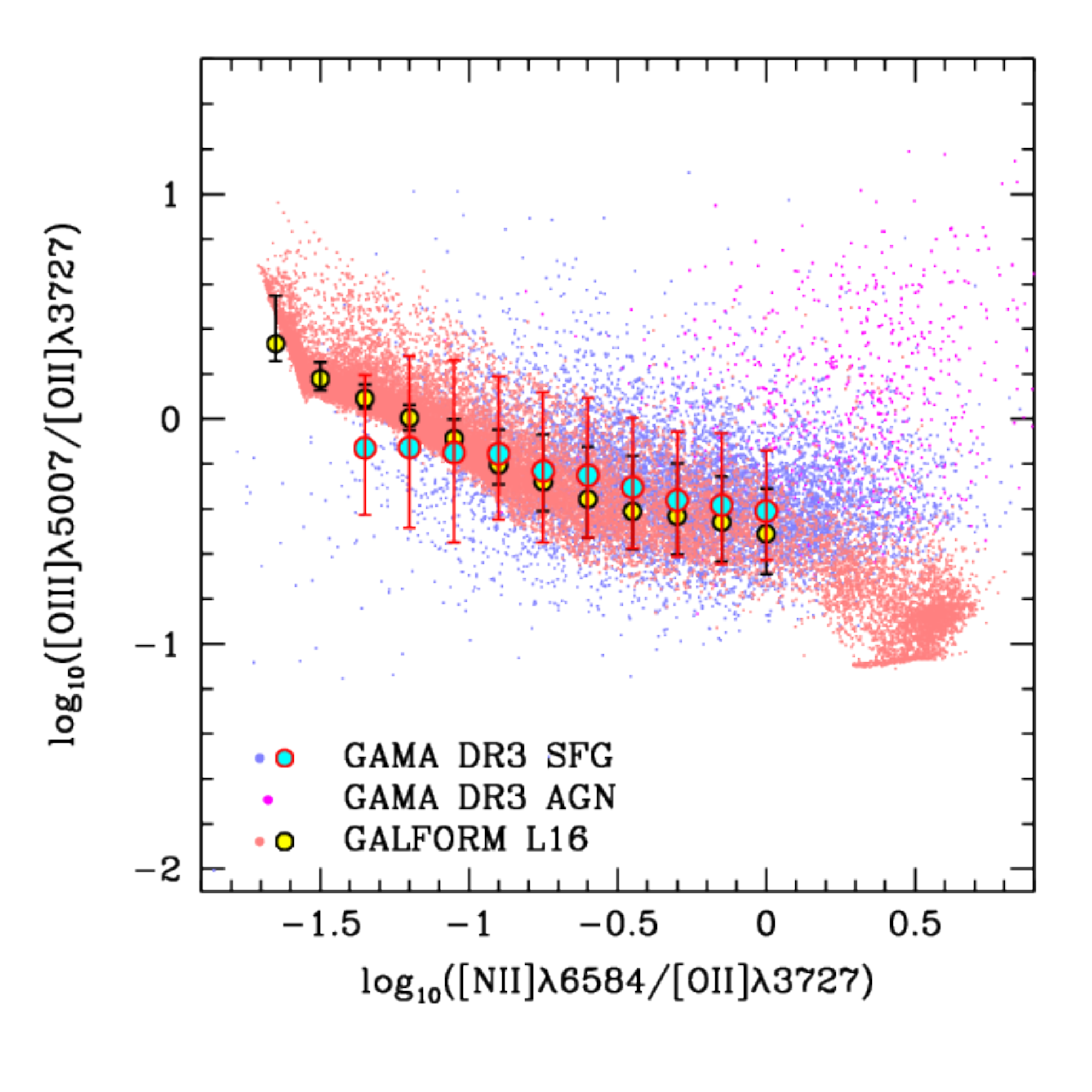}
        	\caption{Other line ratio diagrams at low redshift, comparing the model predictions with those from the GAMA survey. The symbols follow those in Fig.~\ref{fig:bpt_z0.131}. The top panel shows [OIII]/H$\beta$ plotted against [SII]/H$\alpha$ and the lower panel shows [OIII]/[OII] against [NII]/[OII]. In the event these lines are part of a doublet, the wavelength of the line used is included in the label. }
    \label{fig:otherlines}
\end{figure}

Here we show examples of BPT diagrams which feature other emission line ratios. For the samples considered in the main paper, these other lines are only available for the low redshift data from the GAMA survey. Since the BPT locus occupied by the galaxies depends heavily on the survey selection, we therefore restrict this comparison to low redshift. 
The top panel of Fig.~\ref{fig:otherlines} replaces [NII] with [SII] in the denominator of the line ratio plotted on the $x$-axis, following the examples of other line diagnostic diagrams presented in \cite{Byler:2017}. The GAMA objects are classified as being star-forming (SFGs) or AGN using the classic BPT diagram (Fig.~\ref{fig:bpt_z0.131}). 
The top panel of Fig.~\ref{fig:otherlines} shows that when using [SII]/H$\alpha$ instead of [NII]/H$\alpha$, SFGs and AGNs are no longer separated. There is reasonable overlap between the model predictions and the observations, with the comparison affected by the selection applied to the model, which allows lower values of [OIII]/H$\beta$ than are seen in the data. In the version of the BPT diagram plotted in the lower panel of Fig.~\\ref{fig:otherlines}, there is again considerable overlap between SFGs and AGNs. The model predictions are again in reasonable agreement with the observations, although the predicted [OIII]/[OII] ratio declines somewhat more steeply with increasing [NII]/[OII] than is observed.

\bsp	
\label{lastpage}
\end{document}